\documentclass[%
superscriptaddress,
 reprint,
nofootinbib,
 amsmath,amssymb,
 aps,multicol
]{revtex4-2}

\usepackage{graphicx}% Include figure files
\usepackage{dcolumn}% Align table columns on decimal point
\usepackage{bm}% bold math
\usepackage{xcolor}

\begin{document}

\preprint{}

\title{Constraints on Gluon Distribution Functions in the Nucleon and Nucleus from Open Charm Hadron Production at the Electron-Ion Collider}

\author{Matthew Kelsey}
\email{mkelsey@wayne.edu}
\affiliation{Wayne State University, Detroit, MI 48202, USA}
\affiliation{Lawrence Berkeley National Laboratory, Berkeley, CA 94720, USA}

\author{Reynier Cruz-Torres}
\affiliation{Lawrence Berkeley National Laboratory, Berkeley, CA 94720, USA}

\author{Xin Dong}
\affiliation{Lawrence Berkeley National Laboratory, Berkeley, CA 94720, USA}

\author{Yuanjing Ji}
\affiliation{Lawrence Berkeley National Laboratory, Berkeley, CA 94720, USA}

\author{Sooraj Radhakrishnan}
\affiliation{Kent State University, Kent, OH 44242, USA}
\affiliation{Lawrence Berkeley National Laboratory, Berkeley, CA 94720, USA}

\author{Ernst Sichtermann}
\affiliation{Lawrence Berkeley National Laboratory, Berkeley, CA 94720, USA}

%\author{Nu Xu}
%\affiliation{Lawrence Berkeley National Laboratory, Berkeley, CA 94720, USA}

\date{\today}

\begin{abstract}
The Electron-Ion Collider (EIC) at Brookhaven National Laboratory will be a precision Quantum Chromodynamics machine that will enable a vast physics program with electron+proton/ion collisions across a broad center-of-mass range. Measurements of hard probes such as heavy flavor in deep inelastic scatterings will be an essential component to the EIC physics program and are one of the detector R\&D driving aspects. In this paper we study the projected statistical precision of open charm hadron production through exclusive hadronic channel reconstruction with a silicon detector concept currently being developed using a PYTHIA-based simulation. We further study the impact of possible intrinsic charm in the proton on projected data, and estimate the constraint on the nuclear gluon parton distribution function (PDF) from the charm structure functions $F_{2}^{c\overline{c}}$ in $e$+Au collisions using a Bayesian PDF re-weighting technique. Our studies show the EIC will be capable delivering an unprecedented measurement of charm hadron production across a broad kinematic region and will provide strong constraints to both intrinsic charm and nuclear gluon PDFs.      

\end{abstract}

\maketitle
%\tableofcontents
\section{Introduction}

The Electron-Ion Collider (EIC), a US-based facility planned to be constructed at the current Relativistic Heavy-Ion Collider (RHIC) facility at Brookhaven National Laboratory, will be a state-of-the-art Quantum Chromodynamics (QCD) laboratory~\cite{Accardi:2012qut,Aschenauer:2017jsk,AbdulKhalek:2021gbh}. The EIC will be capable of delivering high-luminosity ($10^{34}$ cm$^{-2}$ s$^{-1}$) collisions of electrons on (polarized) protons, light and heavy ions over a large span of center-of-mass energies ranging from $\sqrt{s}$ = 20 to 141 GeV~\cite{eRHIC:preCDR}. Some of the flagship EIC measurements include studies of the spin structure of protons and light ions, the partonic structure of light and heavy ions, partonic transport in nuclear matter, and hadronization.

\begin{figure}[htb]
    \centering
    \includegraphics[width=0.4\textwidth, height=6.5cm]{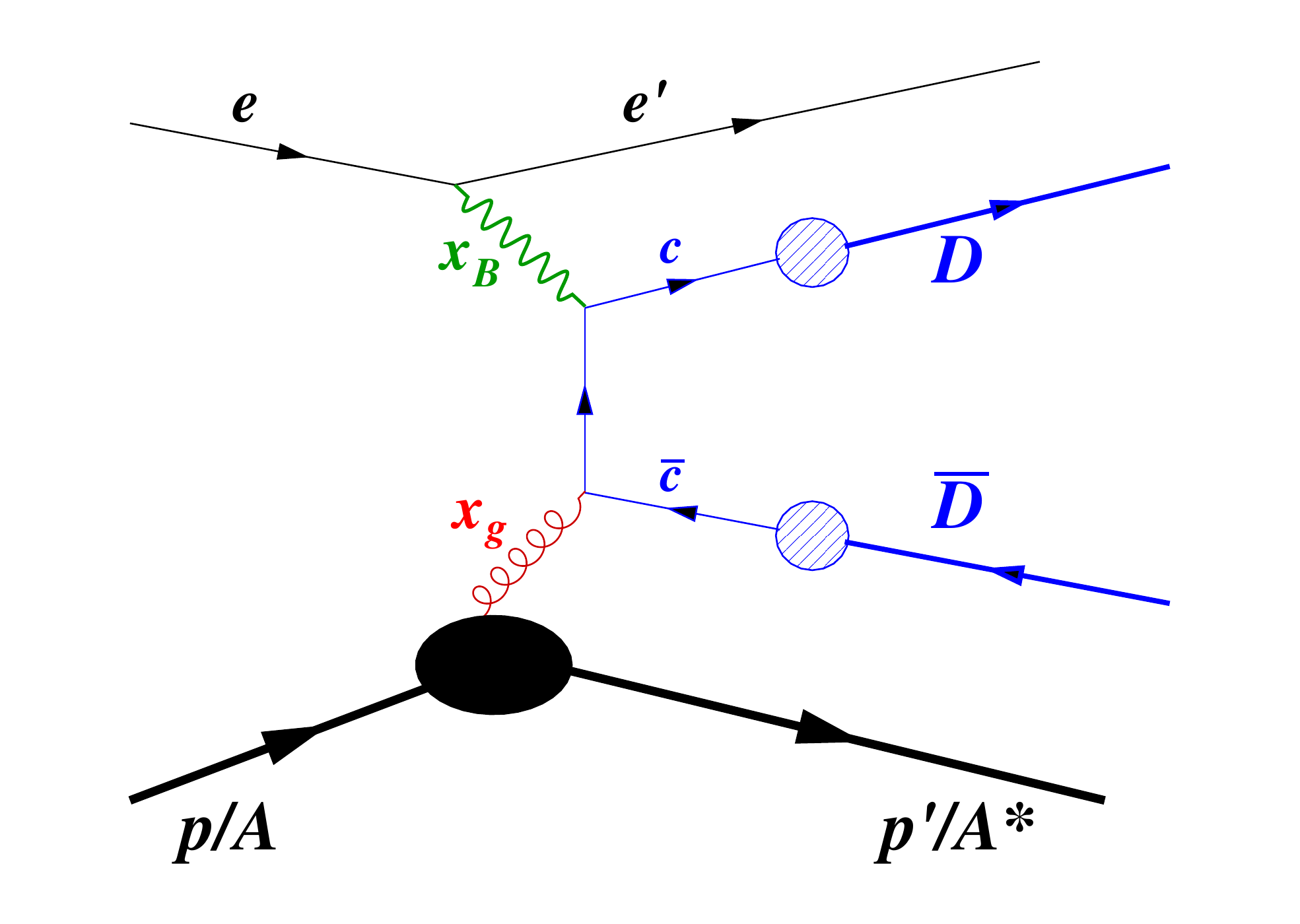}
    \caption{Leading-order diagram for charm and anti-charm pair production in $e$+$p$/ion deep inelastic scatterings (DIS).}
    \label{fig:diagram}
\end{figure}

Measurements of heavy flavor hadrons (hadrons containing a charm or bottom quark) in deep inelastic scatterings (DIS) are valuable since at leading-order (LO) heavy quark pairs are produced via photon-gluon fusion as shown in Fig.~\ref{fig:diagram}, and have direct and clean access to the gluonic structure of nucleons/nuclei. (Nuclear) parton distribution functions (PDF) are an essential ingredient in understanding measurements of nuclear collisions and are of broad interest in the particle and nuclear physics communities. 

\begin{figure}[!htb]
    \centering
    \includegraphics[width=0.48\textwidth]{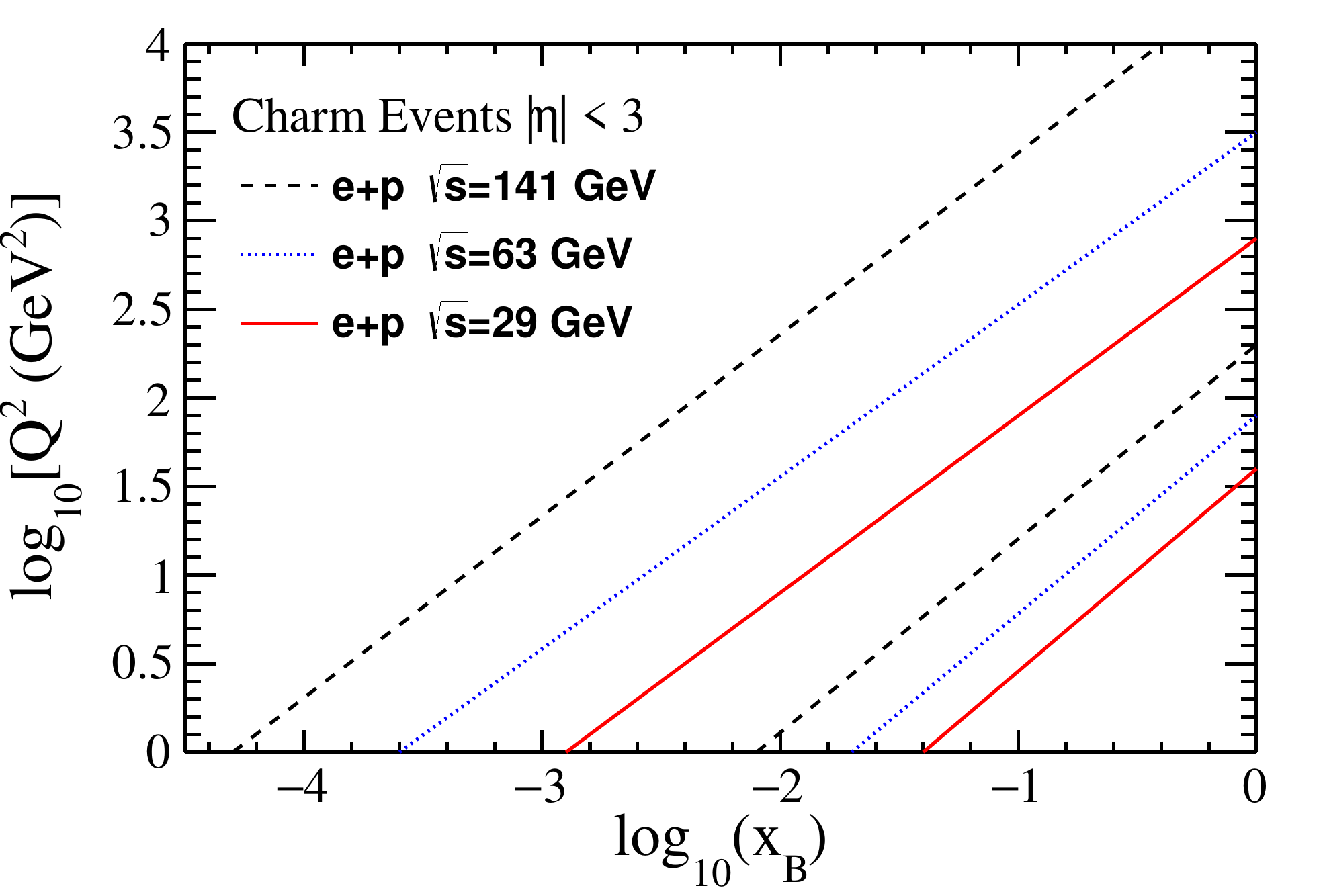}
    \caption{$Q^{2}$ vs. $x_B$ coverage of charm events with at least one charm hadron with $|\eta|<3$ for three beam-energy configurations.}
    \label{fig:cq2x}
\end{figure}

Previous generations of DIS experiments have measured the reduced charm cross-sections in lepton-proton collisions over a limited kinematic range with mostly $x_{B}<0.1$, with $x_{B}$ defined as the Bjorken scaling variable, and have been well described by perturbative QCD (pQCD)~\cite{Abramowicz2018}. At the EIC, kinematic coverage of measurements of charm hadrons will extend to $x_{B}>0.1$ as shown in Fig.~\ref{fig:cq2x} for $e+p$ collisions (and also to very low $x_{B}<10^{-4}$ for higher beam energies). Due to this coverage, measurements of charm hadron production will be able to constrain PDFs at large parton longitudinal momentum fraction, $x_{p}$. This will also provide information to elucidate the EMC effect and short range correlations from $e+$A collisions at large $x_{p}$~\cite{Malace:2014uea}. Compared to inclusive and semi-inclusive light hadron data, heavy flavor production will shed light on the understanding of the gluonic role to the short range correlations in this region. 

The broad kinematic coverage will also be interesting as it probes the potential intrinsic charm (IC) contribution to the proton wave function, which was first proposed in the paper by Brodsky et al.~\cite{Brodsky:1980pb}. Intrinsic heavy quarks, referring to a heavy $Q\overline{Q}$ pair which couples to the valance quarks in the proton forming a five quark Fock state $|uudQ\overline{Q}\rangle$, have been of considerable interest for several decades and are not experimentally confirmed (see e.g. this review~\cite{BRODSKY2017108} and references therein). Separate from the "extrinsic" charm component, arising from perturbative gluon splittings, the IC component is non-perturbartive in nature and can't be calculated with pQCD. Therefore, an understanding of the IC PDF which will be important for precise and accurate (n)PDF determinations from global fits, must come from experimental data. Most analyses find an IC component at large values of $x_{p}$ stemming from the large charm quark mass~\cite{PhysRevD.99.116019,BRODSKY2017108,Brodsky:1980pb,Hou2018,Martin:2009iq,Jimenez-Delgado:2014zga} (also for intrinsic bottom~\cite{Lyonnet2015}). This limits the constraints from existing DIS data. More recent studies of data from the Large Hadron Collider (LHC) have been able to place limits on IC~\cite{Bednyakov2019}. There recently have been additional proposed measurements to probe IC in hadron collisions at RHIC and LHC ~\cite{GONCALVES201059}, and SeaQuest~\cite{PhysRevC.103.035204}, which will provide valuable insight into IC before the EIC era. However, as we show in this Paper, charm hadron production at the EIC will be well-positioned to study IC with high precision over a broader kinematic range with respect to previous DIS experiments.

Charm hadron production measurements at an EIC have been explored in previous studies~\cite{Aschenauer:2017oxs,Chudakov:2016ytj,Wong:2020xtc,wong2020proposed,PhysRevD.103.074023}. In Ref.~\cite{Aschenauer:2017oxs} charm events were tagged by secondary charged kaon tracks while Ref.~\cite{Chudakov:2016ytj} reported charm mesons decays with secondary vertex displacement. In Ref.~\cite{PhysRevD.103.074023} charm jets in charge current DIS are studied by utilizing a displaced track counting method. Recently there has been a rapid development of detector R\&D for the EIC including precision tracking detectors~\cite{arrington2021eic,wong2020proposed}. In this Paper we report the statistical projections of charm production in $e+p$/Au collisions with exclusive hadronic decays utilizing the secondary vertexing capabilities provided from a silicon vertex detector aimed for future EIC experiments. We additionally report the impact on these measurements from intrinsic charm and the constraints these measurements will have on nuclear gluon PDFs.

In this analysis, we haven't considered  the final state effects on the charm hadron production, e.g. the possible cold nuclear medium impact on the charm quark energy loss and hadronization, which are largely uncertain at the present stage. We envision that future EIC experiments with significant statistics over different collision energies and nuclei species will allow us to conduct extensive measurements on various differential charm hadron production, e.g. anti-charm-to-charm ratio, charm hadrochemistry, and charm fragmentation in jets, to offer insights towards the understanding of these final state effects.

This Paper is organized as follows: Section~\ref{sec:s2} is dedicated to the simulation setup used to study charm reconstruction with an EIC detector; In Section~\ref{sec:s3} we will provide the projections for the charm hadron reduced cross-sections and structure functions $F_{2}^{c\overline{c}}$ with the expected integrated luminosities; In Section~\ref{sec:s4} we study the impact of an IC component in the proton on $F_{2}^{c\overline{c}}$ from more recent global PDF fits; In  Section~\ref{sec:s5} we will show the constraint on the nuclear gluon PDF from the projected $F_{2}^{c\overline{c}}$; In Section~\ref{sec:s6} we will summarize our conclusions. We note here that in Sections~\ref{sec:s4} and~\ref{sec:s5} the PDF sets used were generated from the LHAPDF 6 library~\cite{Buckley2015}.     

\begin{figure}[!htb]
    \centering
    \includegraphics[width=0.48\textwidth]{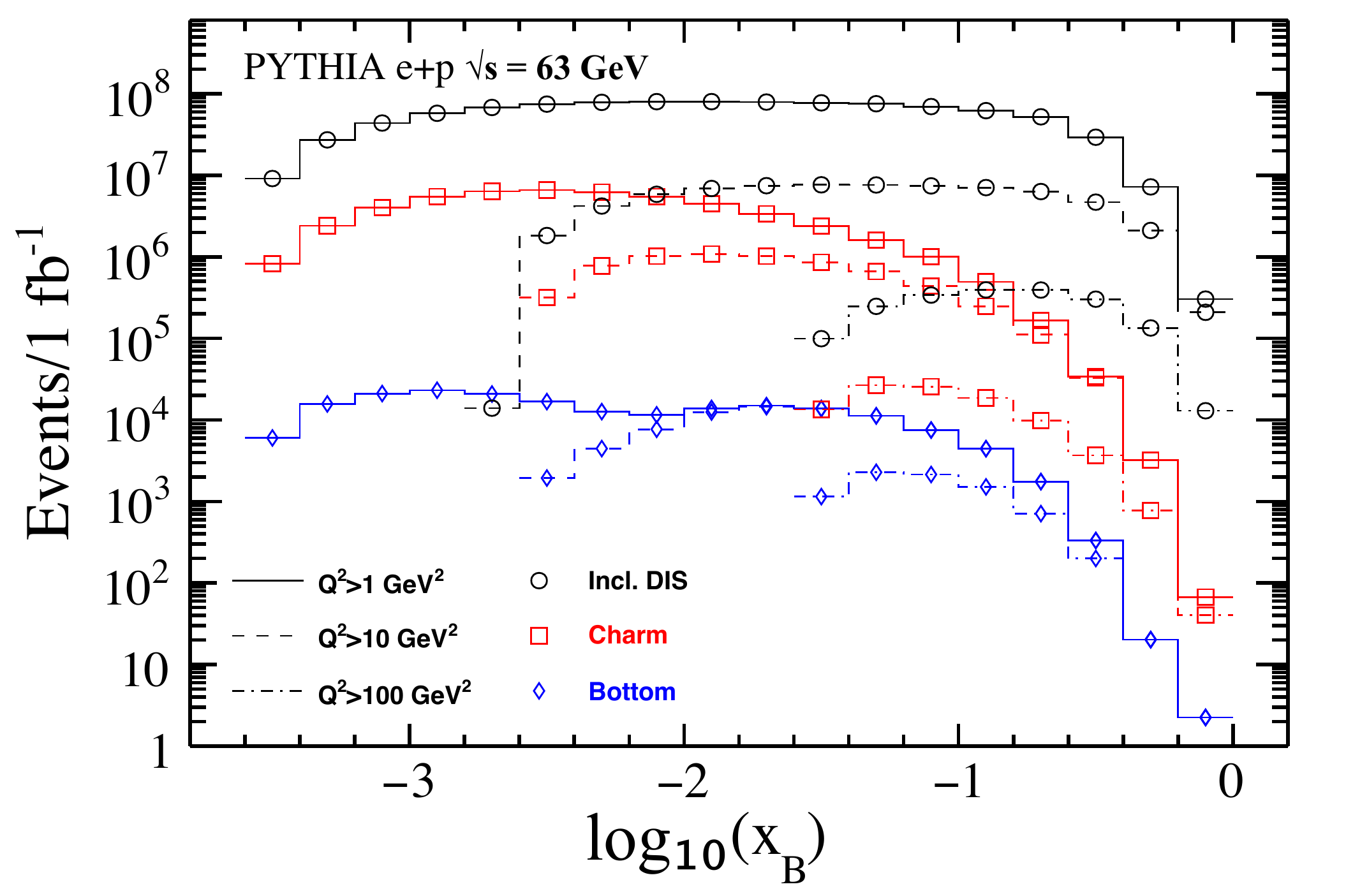}
    \caption{Event yields per 1 fb$^{-1}$ integrated luminosity in e+p $\sqrt{s}=63$ GeV PYTHIA collisions for inclusive DIS (open black circles), charm (open red squares), and bottom (open blue diamonds), for three lower bounds of $Q^{2}$ as depicted by the various line styles.}
    \label{fig:yields}
\end{figure}

\section{Simulation Setup}\label{sec:s2}

For the studies reported in this Paper we utilize the PYTHIA 6.4 event generator~\cite{Sj_strand_2006} with the settings outlined in~\cite{PYTHIA6} to generate $e$+$p$ collisions. Events are generated with vector-meson diffractive and resolved processes, semi-hard QCD 2$\rightarrow$2 scattering, neutral boson scatterings off heavy quarks within the proton, and photon-gluon fusion. The latter two processes are predominantly responsible for heavy flavor quark production.

The kinematic variables used through this paper are defined as follows: In the one-photon-exchange approximation, an incoming electron of four momentum $e$ scatters into a final state $e'$ via the emission of a virtual photon of four momentum $q=e-e'$, which subsequently interacts with the hadron beam with four momentum $p$. We follow the convention that the hadron beam momentum is along the positive $z$ direction. The Bjorken scaling variable is defined as
$x_B \equiv Q^2/(2p \cdot q)$
%, where $p^{\mu}_2$ is the target four momentum,
and $Q^2\equiv -q^2$ is minus the square of the four momentum transfer. The inelasticity is defined as $y\equiv p \cdot q /(p \cdot e)$. For the purposes of these studies we do not include any radiative corrections to the incoming/scattered lepton.

\begin{figure}[!htb]
    \centering
    \includegraphics[width=0.45\textwidth]{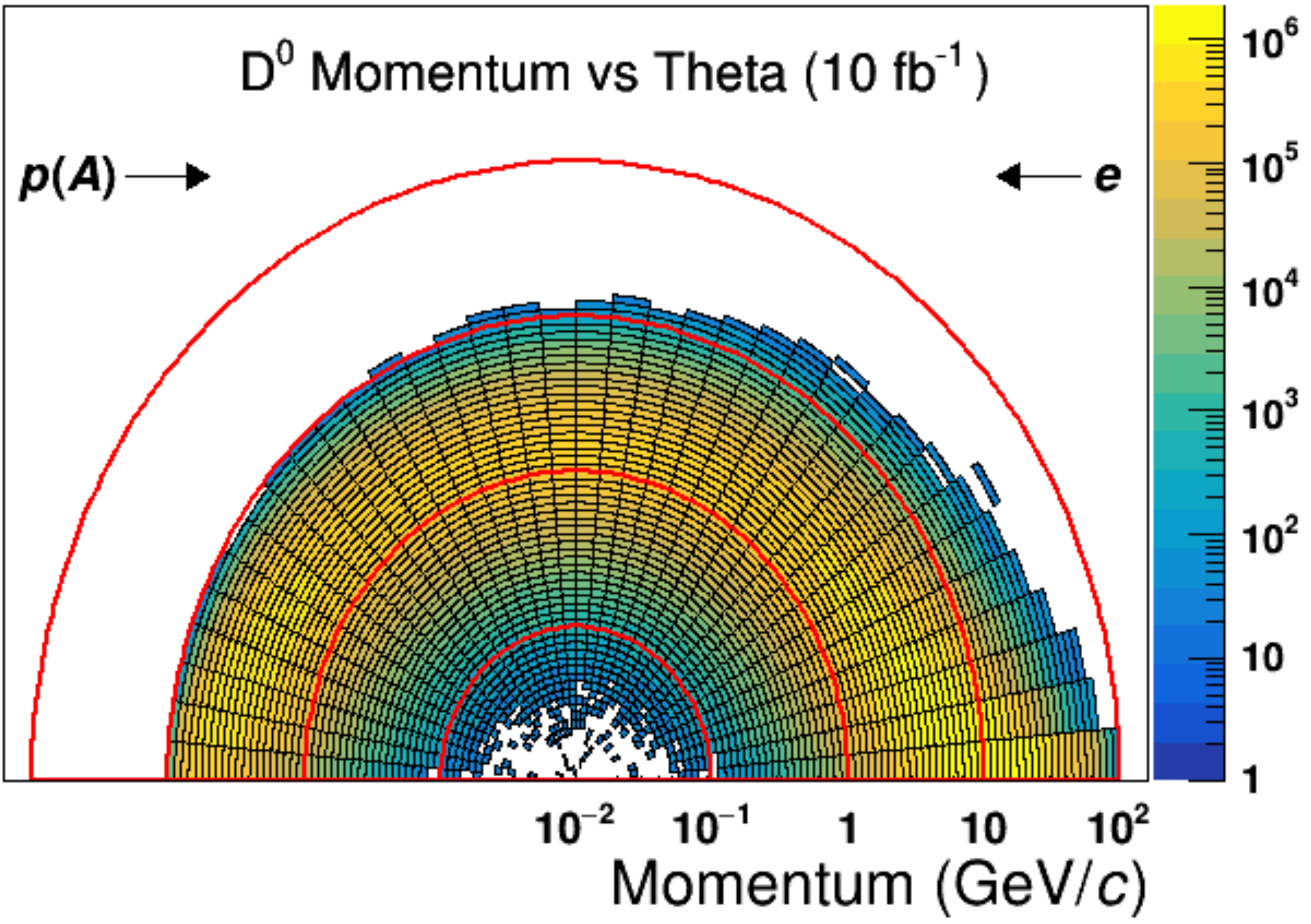}
    \includegraphics[width=0.45\textwidth]{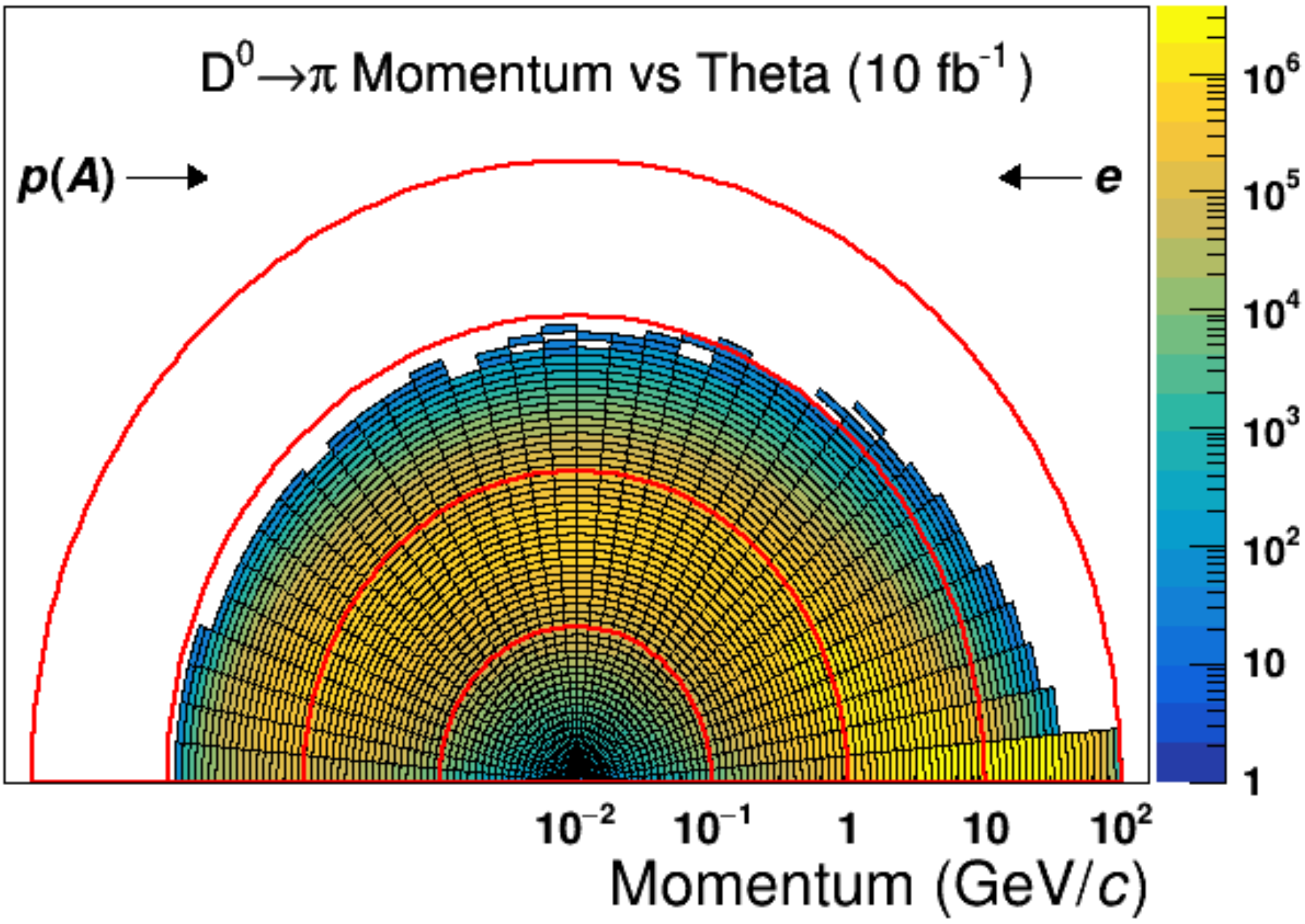}  
    \includegraphics[width=0.45\textwidth]{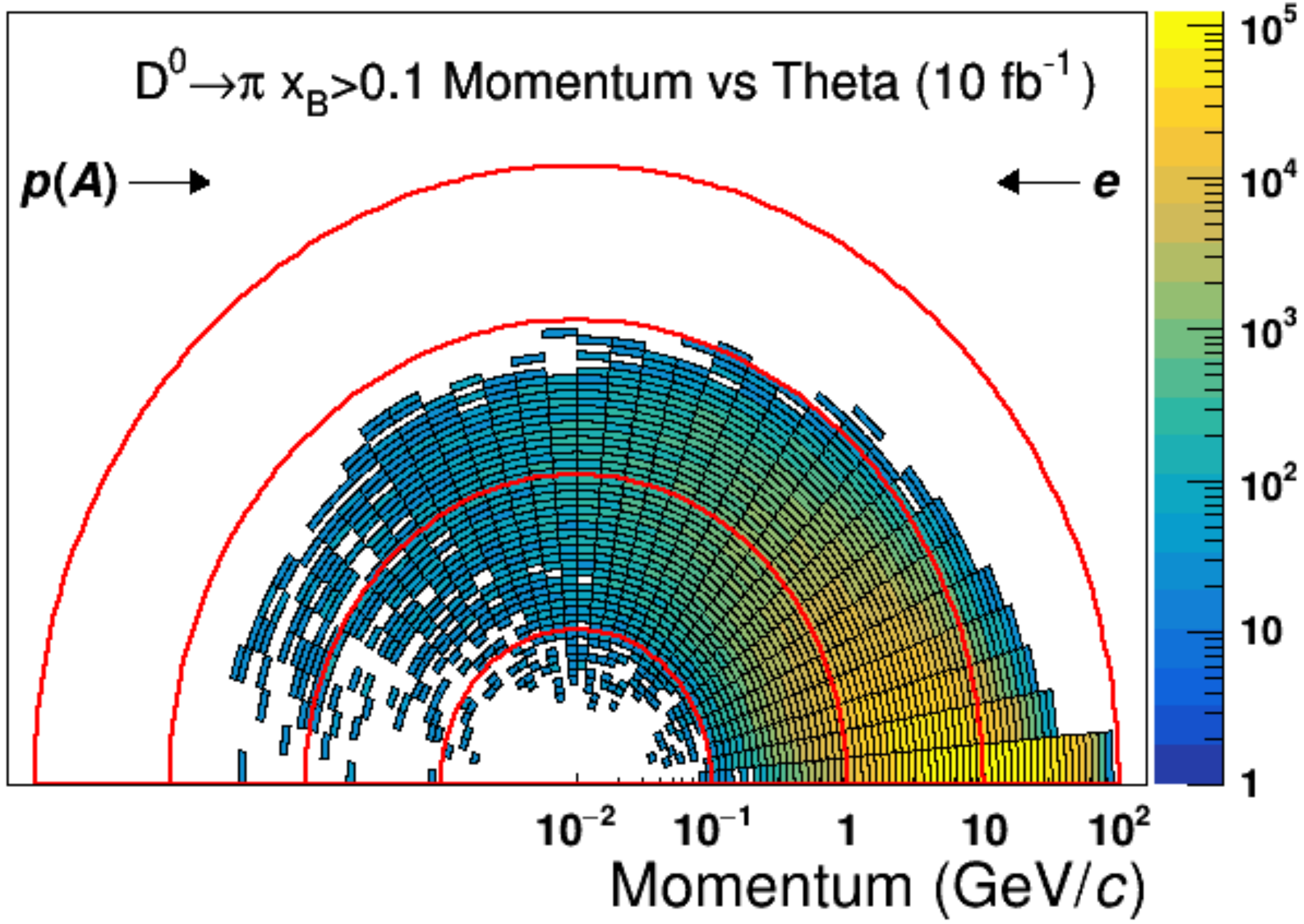}  
    \caption{Kinematic distributions, in polar coordinates, of $D^{0}$ mesons (top) and decay pions (middle and bottom) in $\sqrt{s}=63$ GeV electron-proton collisions generated with PYTHIA 6. Each red semi-circle shows the absolute momentum scale at each order of magnitude as indicated by the $x$-axis intercept. The $z$-axis denotes the yield scaled to 10 fb$^{-1}$. The bottom panel shows the decay-pion distributions after applying an event-level selection of $x_B>0.1$.}
    \label{fig:D0Kin}
\end{figure}

The inclusive yields from PYTHIA with a 10 GeV electron beam colliding with a 100 GeV proton beam ($\sqrt{s}=63$) per 1 fb$^{-1}$ integrated luminosity are shown in Fig.~\ref{fig:yields} for DIS, charm and bottom production. Integrated above $Q^{2}$ = 1 GeV$^{2}$ charm events are about 5.2\% of the total DIS cross section, and bottom events are about 0.02\%. As previously mentioned, for the expected $e$+$p$ beam energy configurations charm hadron production will probe a broad kinematic range in $x_{B}$ and $Q^{2}$ as shown in Fig.~\ref{fig:cq2x} for 18$\times$275 ($\sqrt{s}=141$ GeV), 10$\times$100 ($\sqrt{s}=63$ GeV), and 5$\times$41 ($\sqrt{s}=29$ GeV) GeV $e$+$p$ collisions. Here we place a kinematic cut on the pseudo-rapidity $\eta$ of the charm hadron to be within $\pm$3, which represents the approximate acceptance of an EIC central tracker detector. In Fig.~\ref{fig:D0Kin}, we show the momentum and polar angle distributions for $D^{0}$ (top panel) and $D^{0}\rightarrow\pi$ (middle panel), where the polar angle is defined with respect to the beam axis, for $\sqrt{s}=63$ GeV $e$+$p$ collisions. In the bottom panel we show the decay pion distributions with an event level cut of $x_{B}>0.1$. Other charm hadrons have qualitatively similar distributions. With the current binning the bin edges between the first and second polar sectors at 5 and 175 degrees show approximately $\eta=\pm3$, with positive $\eta$ defined as the hadron beam direction. One can observe in DIS collisions charm hadrons are produced predominantly within $-3<\eta<3$. For events with larger $x_{B}$, the decay hadrons are more populated at large $\eta$. Therefore, the quality of charm hadron measurements at large $x_{B}$ will be highly impacted by the forward tracking system.

Charm hadron reconstruction in the EIC environment is studied using a fast simulation procedure where we smear the particle level momentum and vertex position using parameterized momentum and pointing resolutions. Additionally, a particle identification (PID) scenario, tracking efficiencies, and primary vertex (PV) resolution are included in our fast simulation. The detector concept that guides the tracking and PID performance used here is explained in more detail in Ref.~\cite{arrington2021eic} and references therein. The general design consists of a barrel detector covering $|\eta|<1$ with six silicon pixel layers, and five tapered silicon pixel planes in each the forward and backward regions covering approximately $1<|\eta|<3$. The radial extent of the outer barrel layer and largest forward/backward planes is 43.2 cm. The inner radii of the barrel and forward/backward planes are limited by the beam pipe radius and thickness, which in the nominal interaction region is 3.17 cm and 760 $\mu$m, respectively. At large $|z|$ the beam pipe radius fans out to take into account the finite beam crossing angle, and therefore the inner-most barrel layer and disks have radii of 3.30 cm and 3.20 cm, respectively. We note that we do not explicitly study the effects of a finite beam crossing angle. The inner radii of the detector planes gradually increase at larger $|z|$ to reach a maximum of 5.91 and 4.41 cm for the forward and backward regions, respectively. The total $z$ extent of the tracking system is 242 cm. 

\begin{figure}[htbp]
    \centering
    \includegraphics[width=0.45\textwidth]{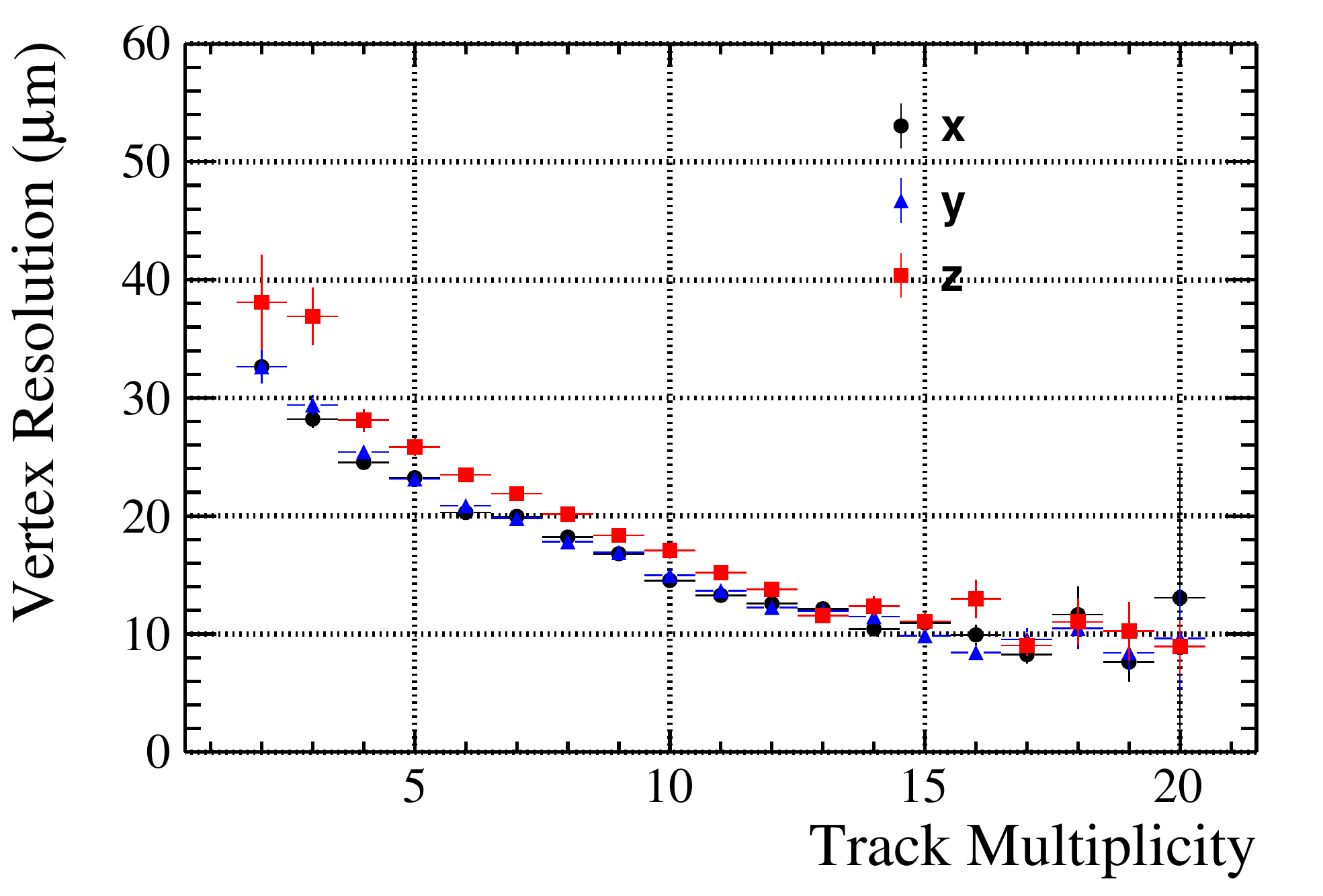}
    \caption{Primary vertex resolution determined by fitting reconstructed tracks with $|\eta|<3$ in the full simulation setup with PYTHIA $e$+$p$ events at $\sqrt{s}=141$ GeV collisions with a uniform 3.0 T magnetic field and event level selection of $Q^{2}>$1 GeV$^{2}$.}
    \label{fig:pv}
\end{figure}

\begin{figure}[htbp]
    \centering
    \includegraphics[width=0.45\textwidth]{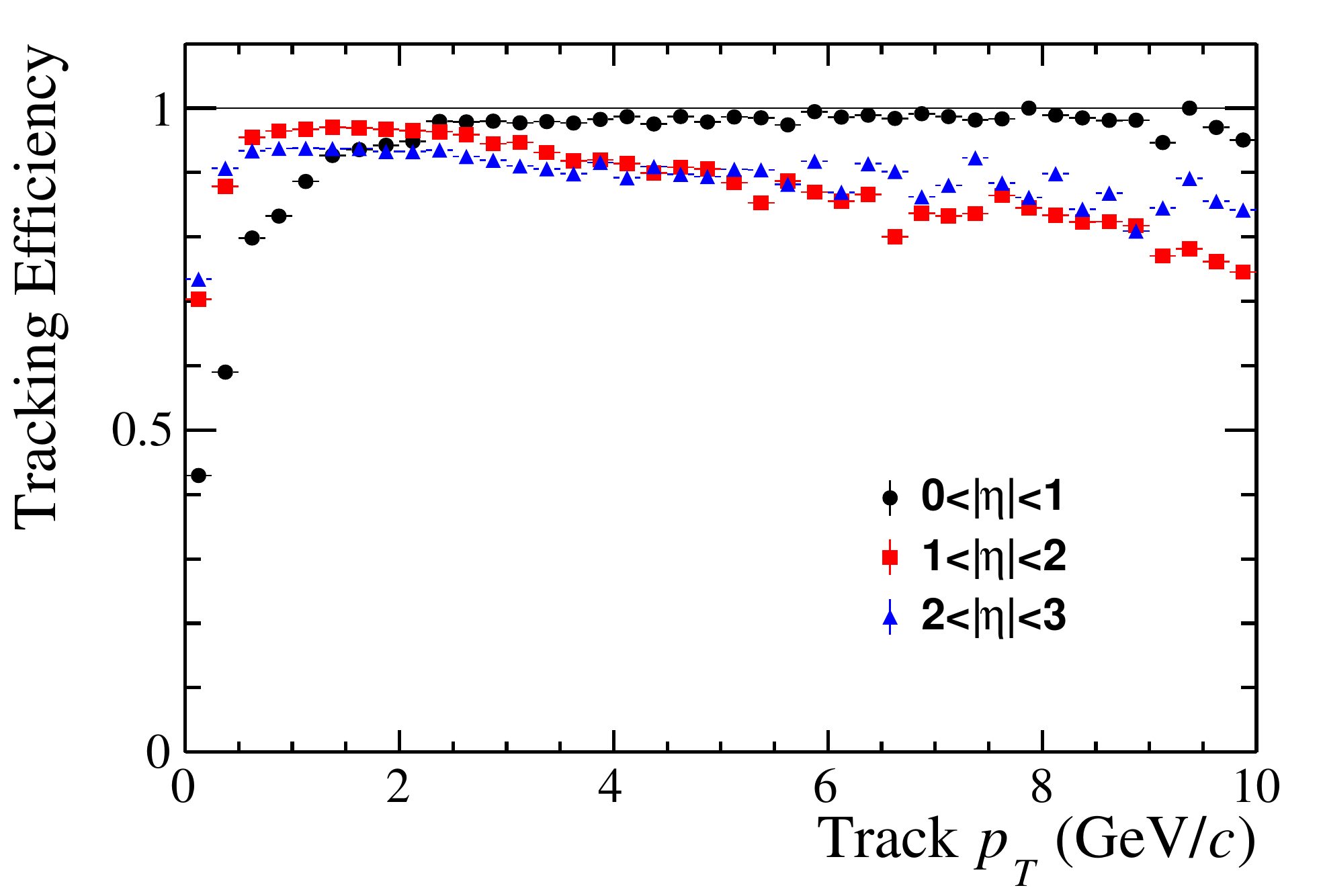}
    \caption{Tracking efficiency determined in the full simulation  in a uniform 3.0 T magnetic field for three different $\eta$ regions.}
    \label{fig:track}
\end{figure}

The relevant momentum and pointing resolution parameterizations, and PID performance used in the fast simulation are tabulated in Table~\ref{tab:sim:smearing} for a uniform longitudinal magnetic field strength of 3 T. We do not consider particles that are produced with $|\eta|>3$. The PV resolution was studied using a GEANT4-based full simulation in the Fun4All framework~\cite{Fun4All,Pinkenburg:2011zza,Pinkenburg:2005zza}. In this study we embed PYTHIA events with an event level selection of $Q^{2}>1$ GeV$^{2}$ and with at least one produced $D^{0}$ in an all-silicon tracker and fit for the PV position using all reconstructed tracks within the detector acceptance ($|\eta|<3$). Fig.~\ref{fig:pv} shows the PV resolution in all three dimensions as a function of track multiplicity and has a resolution around 30 $\mu$m per dimension at very low multiplicity and gradually reduces to about 10 $\mu$m per dimension at a multiplicity greater than fifteen tracks. For charm events the average track multiplicity is nine corresponding to an average PV resolution of around 18 $\mu$m in each dimension. A complete study of the tracking efficiency is not yet available, but as a proxy we use the pion pseudo-tracking efficiency in the same Fun4All simulation using truth level track seeding as shown in Fig.~\ref{fig:track} as a function of track transverse momentum ($p_{T}$) for three different $\eta$ regions. At $|\eta|<1$ we note the low $p_{T}$ efficiency is worse compared to more forward/backward tracks, and is directly related to the minimum $p_{T}$ threshold needed for a track to reach the outer-most barrel layer. For $|\eta|>1$, the efficiency loss at higher $p_{T}$ is due to edges in acceptance at the barrel-to-plane transition region and also the beam pipe openings in the silicon planes at small radii. Particles with $|\eta|<3$ in the fast simulation are randomly removed from an event according to these efficiencies. Only tracks which pass are used to determine a total event multiplicity for the event-by-event multiplicity-dependent PV position smearing.       

\begin{table*}[htb]
\centering
	\caption{Smearing parameters used in fast simulation in different $\eta$ bins: momentum resolution with two sets of magnetic-field configurations, DCA$_{r\phi}$ pointing resolution and particle identification (PID) momentum upper limits. All $p$ and $p_T$ values are in the unit of GeV/$c$.  \label{tab:sim:smearing}}
	\centering
	\begin{tabular}{c | c  c  c}
	$\eta$ & ~~~$\sigma_p/p$ - 3.0~T (\%)~~~  (\%)~~~ & ~~~$\sigma(\rm DCA_{r\phi})$ ($\mu$m)~~~ & ~~~$p_{\rm max}^{\rm PID}$ (GeV/$c$)~~~ \\ \hline 
    ~~~(-3.0,-2.5)~~~ & 0.1$\cdot p$ $\oplus$ 2.0 &  60/$p_T$ $\oplus$ 15 & 10 \\
	(-2.5,-2.0) & 0.02$\cdot p$ $\oplus$ 1.0 &  60/$p_T$ $\oplus$ 15 & 10 \\
    (-2.0,-1.0) & 0.02$\cdot p$ $\oplus$ 1.0 &  40/$p_T$ $\oplus$ 10 & 10 \\
	(-1.0,1.0) & 0.02$\cdot p$ $\oplus$ 0.5 &  30/$p_T$ $\oplus$ 5 & 6\\
	(1.0,2.0) & 0.02$\cdot p$ $\oplus$ 1.0 &  40/$p_T$ $\oplus$ 10 & 50 \\
	(2.0,2.5) & 0.02$\cdot p$ $\oplus$ 1.0 & 60/$p_T$ $\oplus$ 15 & 50 \\
    (2.5,3.0) & 0.1$\cdot p$ $\oplus$ 2.0 & 60/$p_T$ $\oplus$ 15 & 50 \\
	\end{tabular}
\end{table*}

\begin{figure*}[htb]
    \centering
    \includegraphics[width=0.3\textwidth]{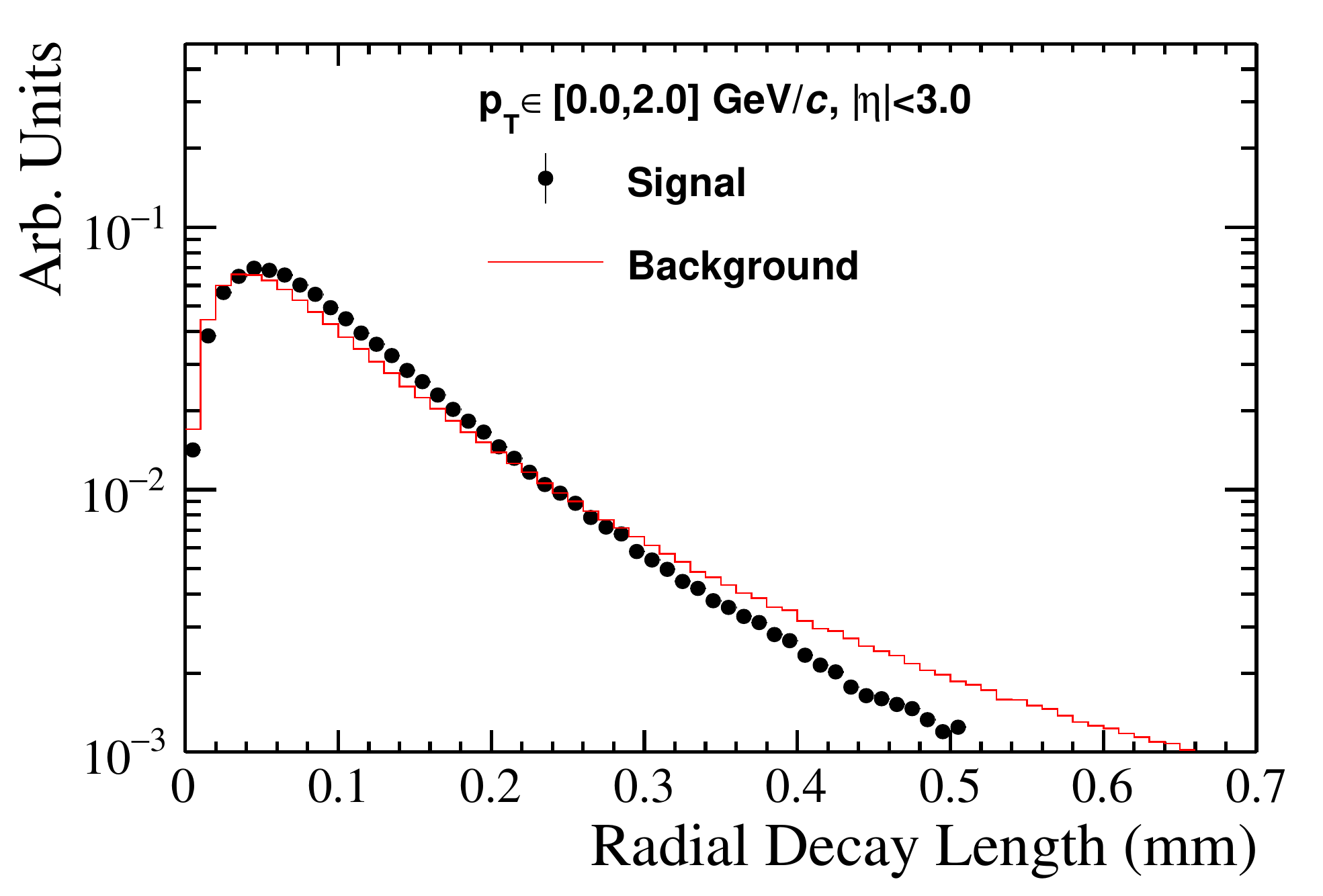}
    \includegraphics[width=0.3\textwidth]{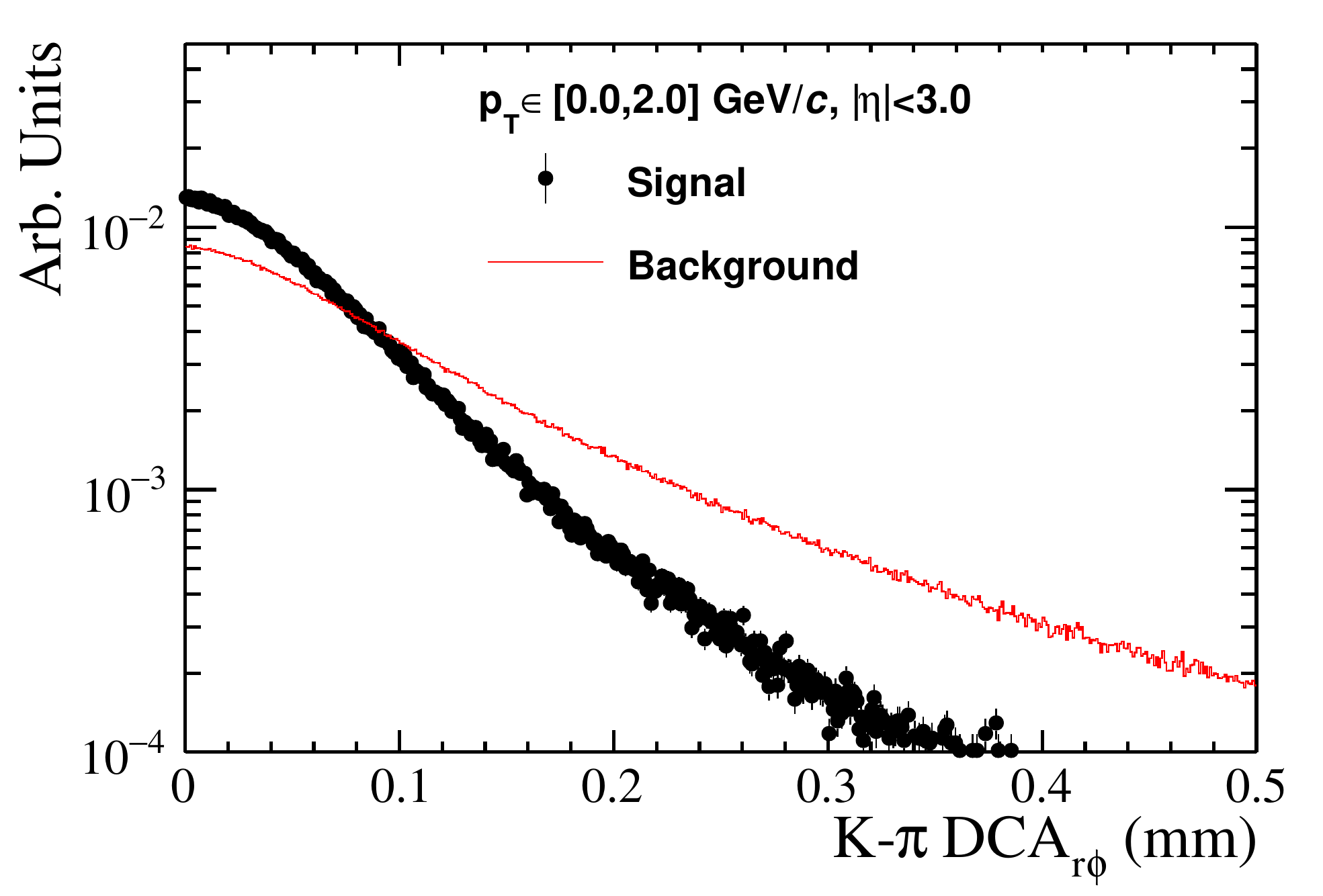}
    \includegraphics[width=0.3\textwidth]{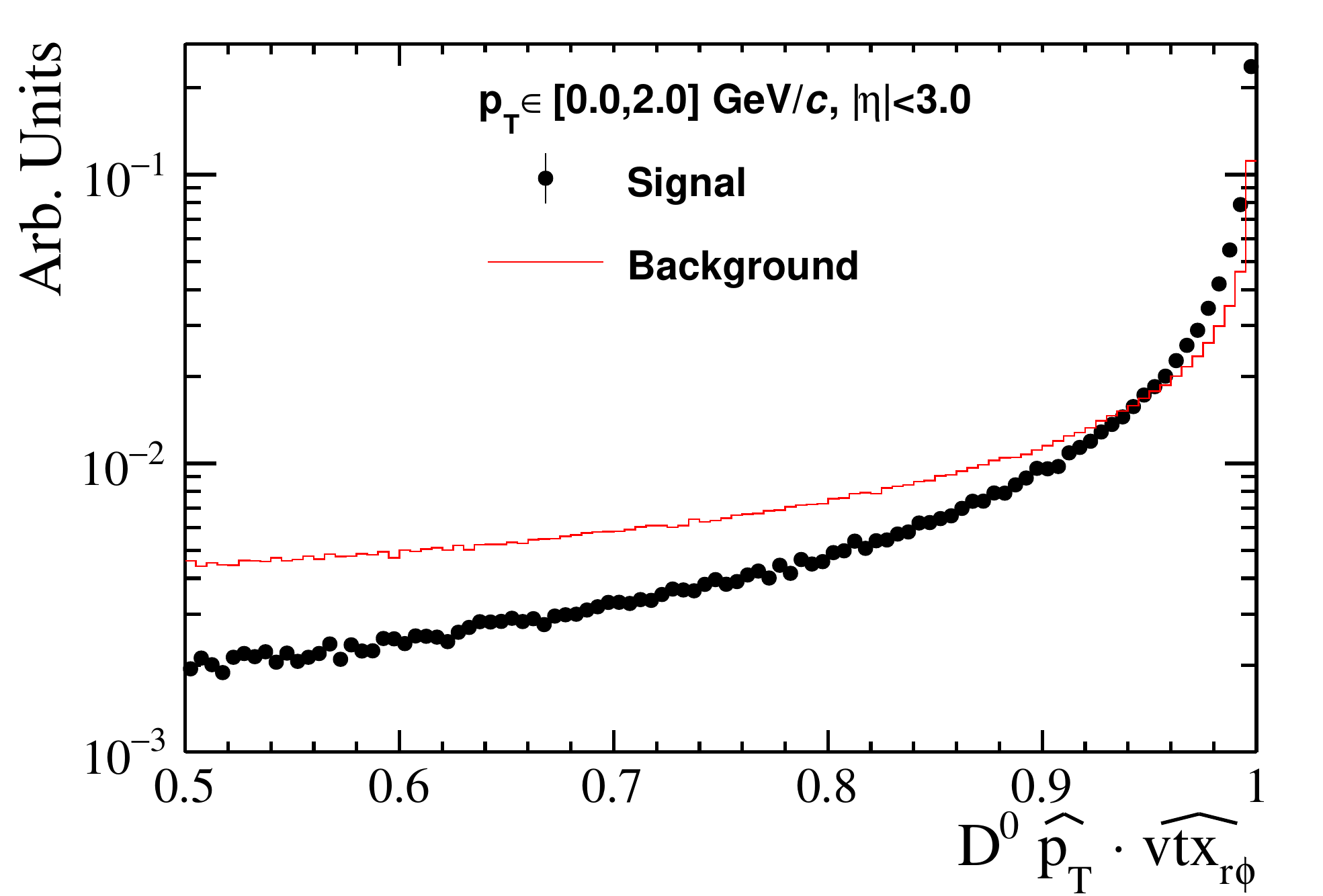}
    \caption{Example fast simulation distributions for the $D^{0}$ radial decay length (left), DCA between kaon and pion tracks (middle), and cosine of the $D^{0}$ radial decay vertex pointing angle (right) for signal (black points) and background (red histograms).}
    \label{fig:topo}
\end{figure*}

Acquiring a high-purity charm hadron data set requires the identification of the secondary decay vertex to reduce combinatorial backgrounds. For these studies we consider only $D^{0}$ mesons, which have a $c\tau$ of approximately 120 $\mu$m~\cite{Zyla:2020zbs}, with the exclusive hadronic $D^{0}\rightarrow K^{-}\pi^{+}$ and charge conjugate decays. The pertinent topological variables are the distance between the $K\pi$ and primary vertices (decay length), the distance-of-closest-approach (DCA) between the $K\pi$ pair, and the cosine of the angle, $\theta$, between the $D^{0}$ momentum and primary-to-secondary vertex vectors. We note here that the pointing resolution is expected to be significantly better in the $r-\phi$ dimensions than the $z$ dimension in the forward and backward rapidity regions, and therefore we only consider topological variables in the transverse plane. In Fig.~\ref{fig:topo} we show all three topological distributions for $D^{0}$ signal (black points) and background (red histogram) with $p_{T}<$ 2 GeV/$c$ and $|\eta|<3$, and an invariant $K\pi$ mass within three sigma of the mass peak. Here both pointing and PV resolutions are folded into the simulation.

\begin{figure*}[htb]
    \centering
    \includegraphics[width=0.3\textwidth]{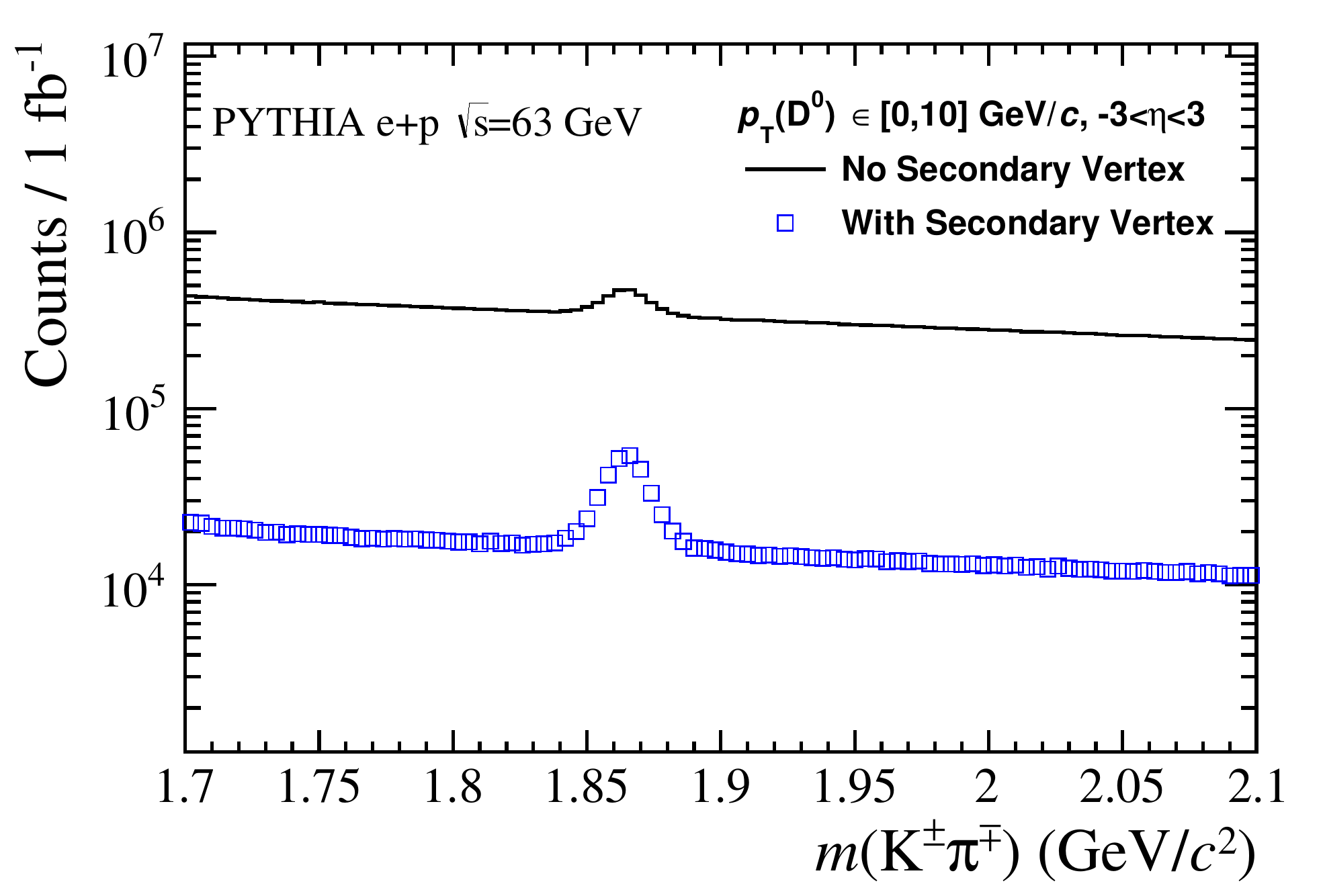}
    \includegraphics[width=0.3\textwidth]{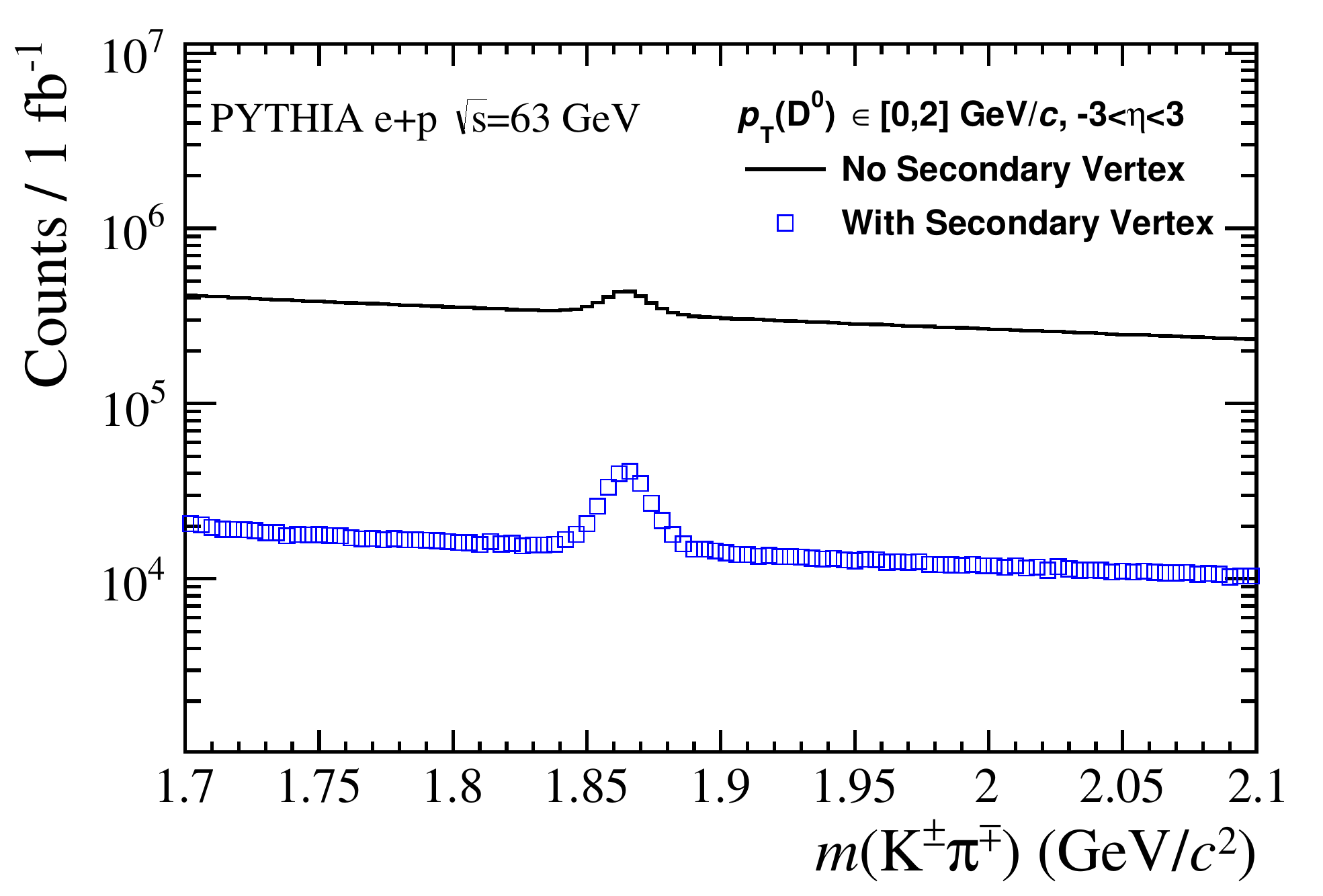}
    \includegraphics[width=0.3\textwidth]{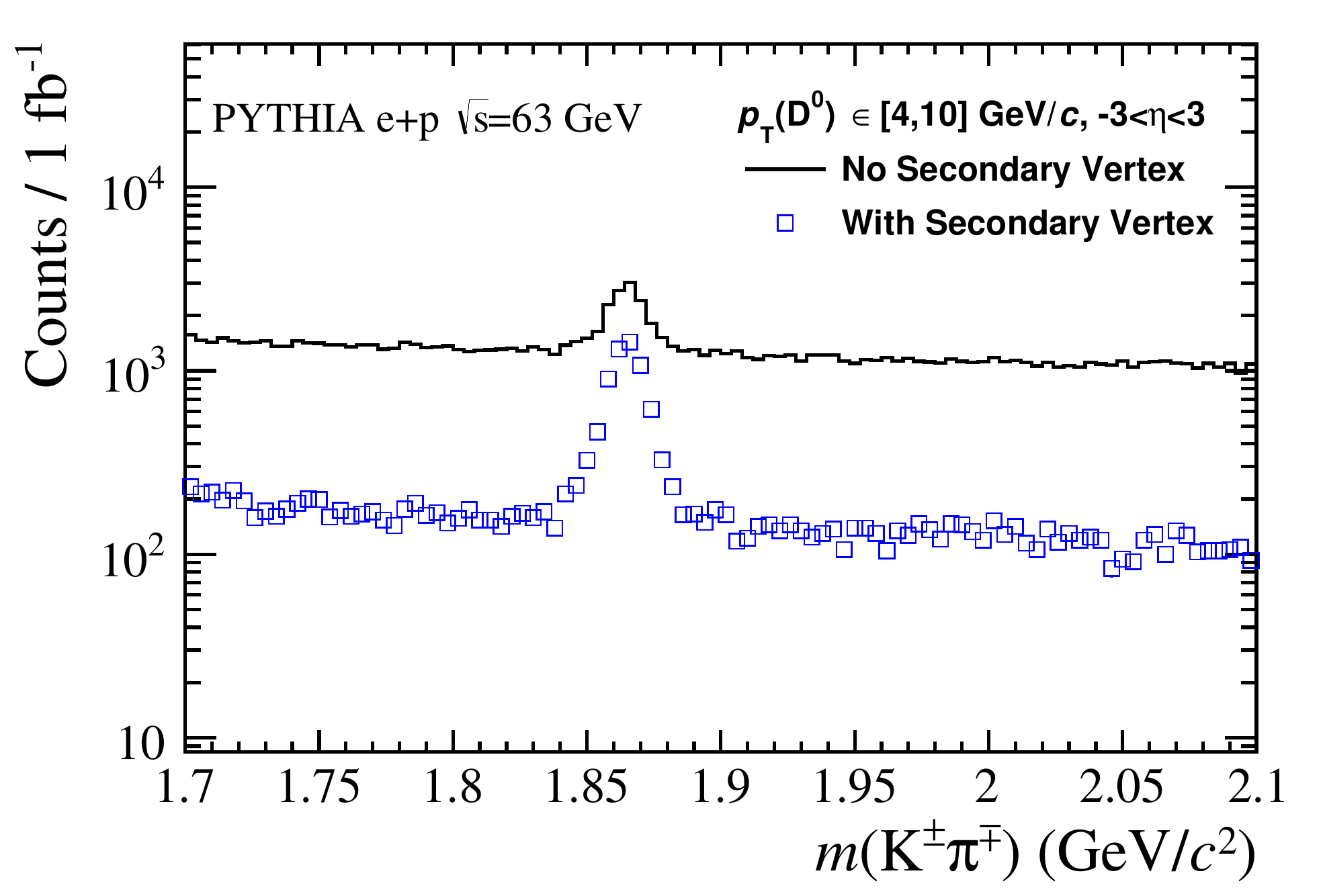}
    \includegraphics[width=0.3\textwidth]{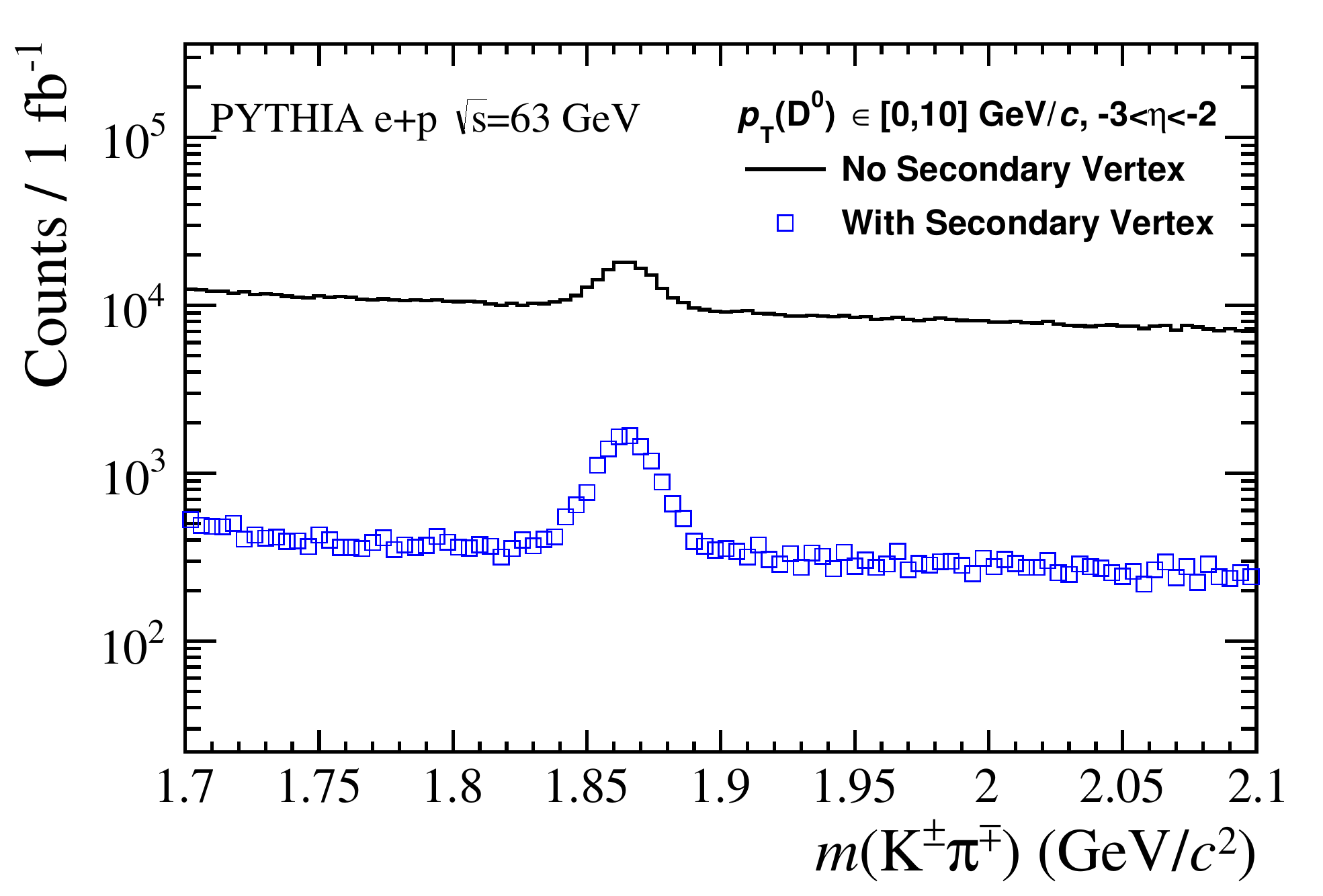}
    \includegraphics[width=0.3\textwidth]{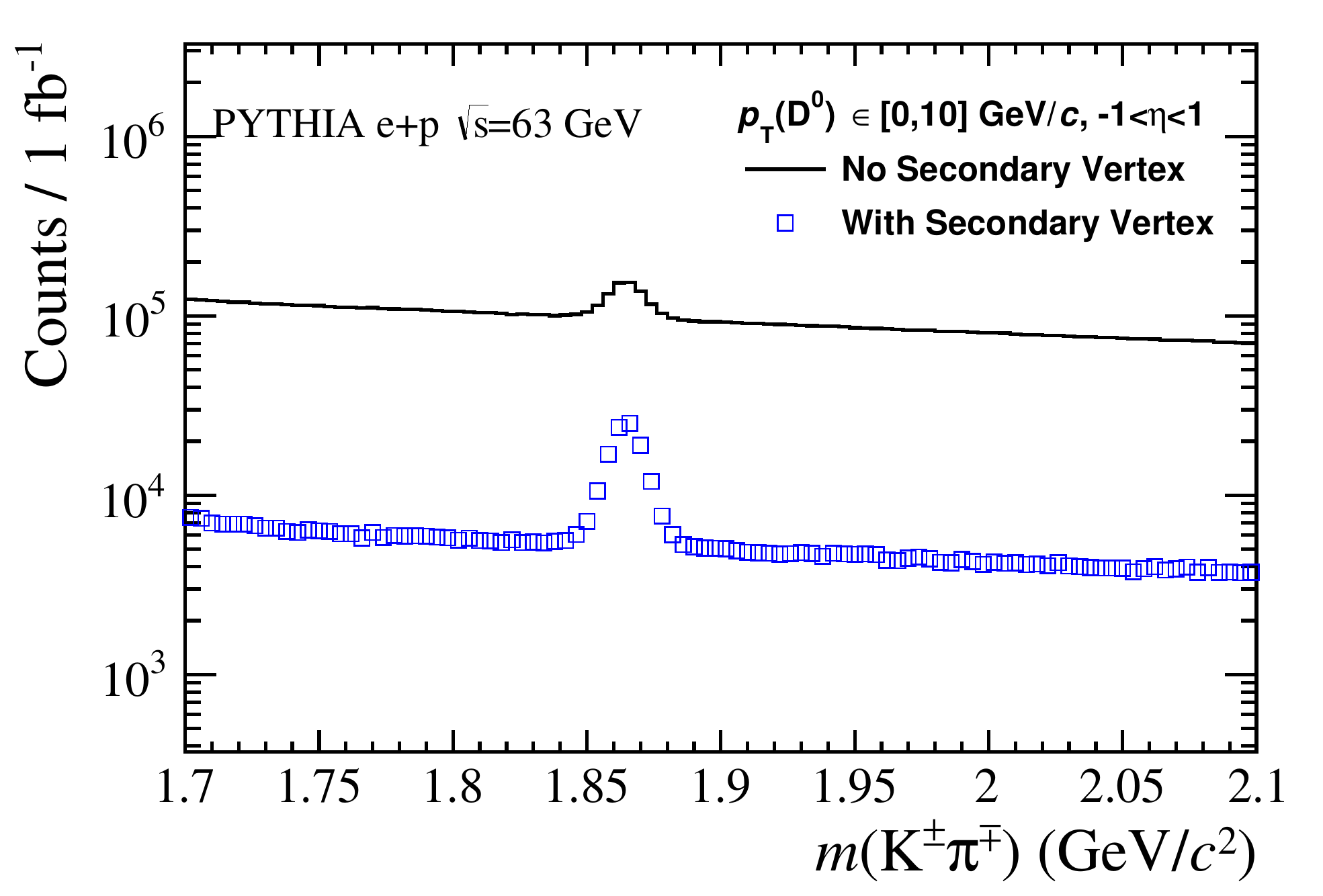}
     \includegraphics[width=0.3\textwidth]{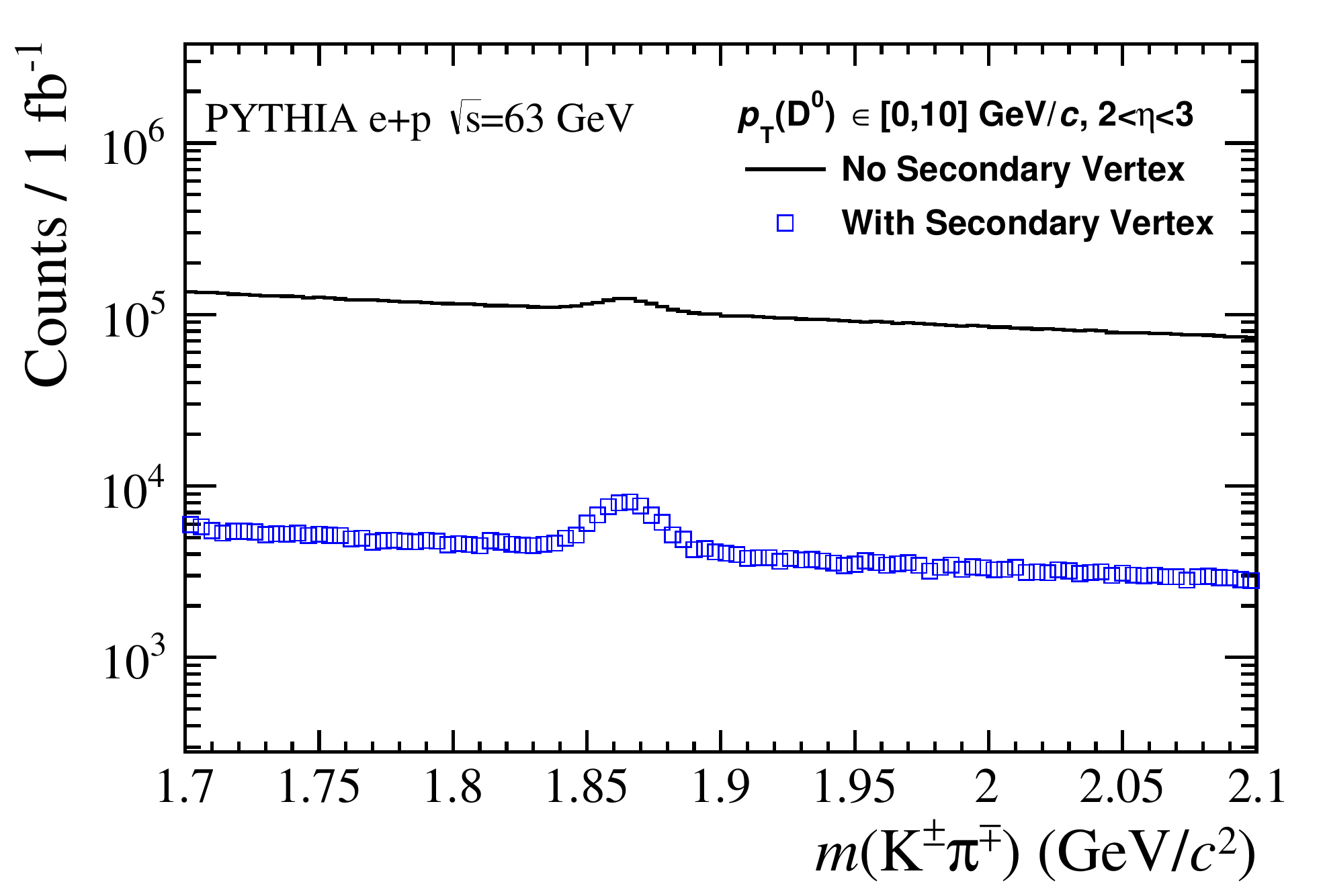}     
    \caption{$D^{0}$ invariant mass distributions from PYTHIA events with (blue squares) and without (black histogram) secondary vertex selection. Each panel shows different $D^{0}$ kinematic regions as specified in the legends.}
    \label{fig:mass}
\end{figure*}

As a baseline selection we utilize the follow requirements to select $D^{0}$ candidates: 1) Radial decay length~$>$ 40 $\mu$m; 2) Pair DCA $<$ 150 $\mu$m; 3) cosine($\theta$) $>$ 0.98. We note a greater signal significance can be achieved with an optimized or more sophisticated selection (e.g., boosted decision tress) and utilizing the longitudinal dimension, and is left to future studies. The impact of these selections on the signal mass peaks in $\sqrt{s}=63$ GeV $e$+$p$ collisions are shown in Fig.~\ref{fig:mass} for various kinematic regions. In general the signal-to-background (S/B) ratio is improved with respect to no secondary vertex selections. Particularly, this improvement is seen at higher values of $p_{T}$. This in effect will provide a data sample with higher signal significance and will reduce systematic uncertainties associated with the signal extraction.

\begin{figure*}[htb]
    \centering
    \includegraphics[width=0.3\textwidth]{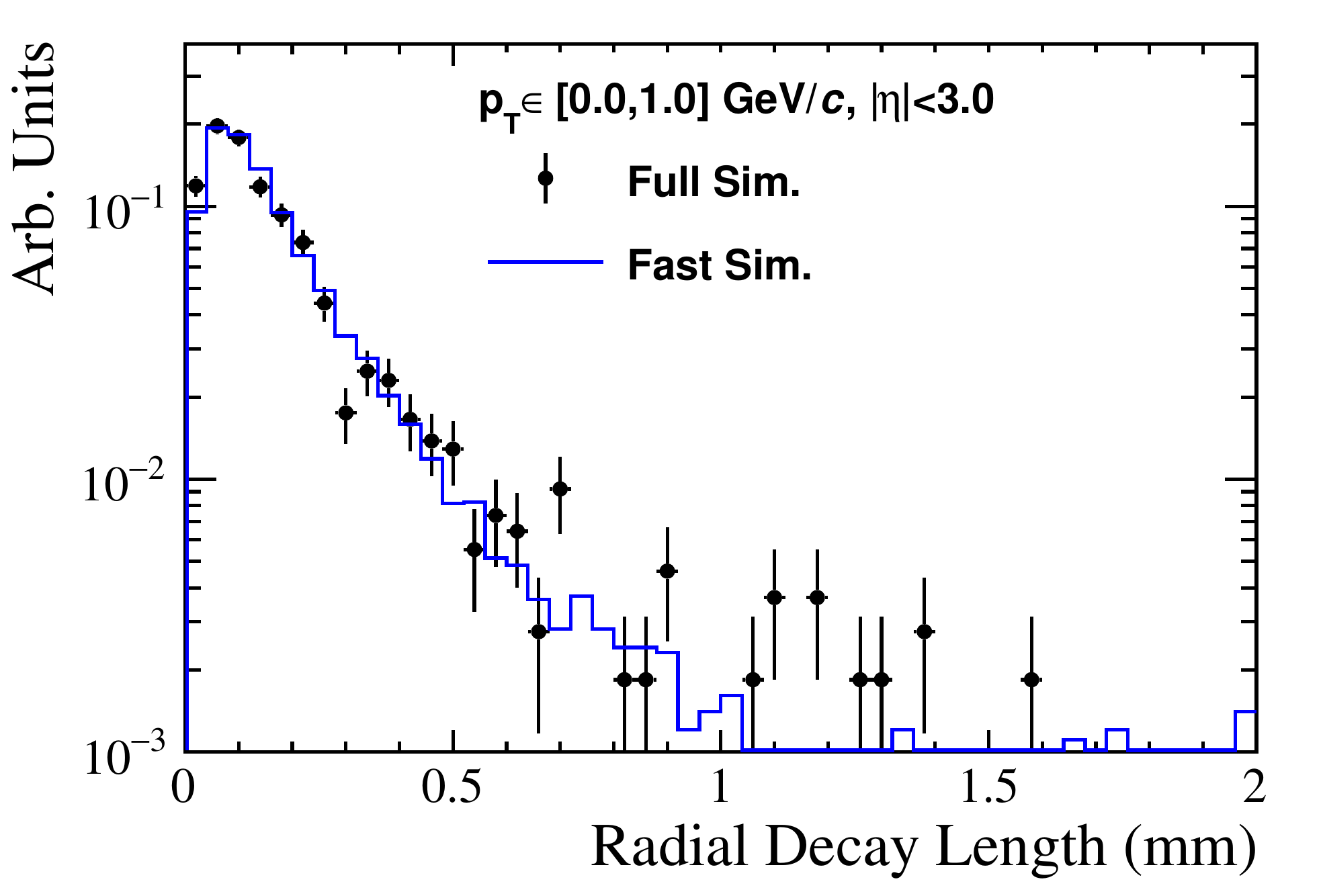}
    \includegraphics[width=0.3\textwidth]{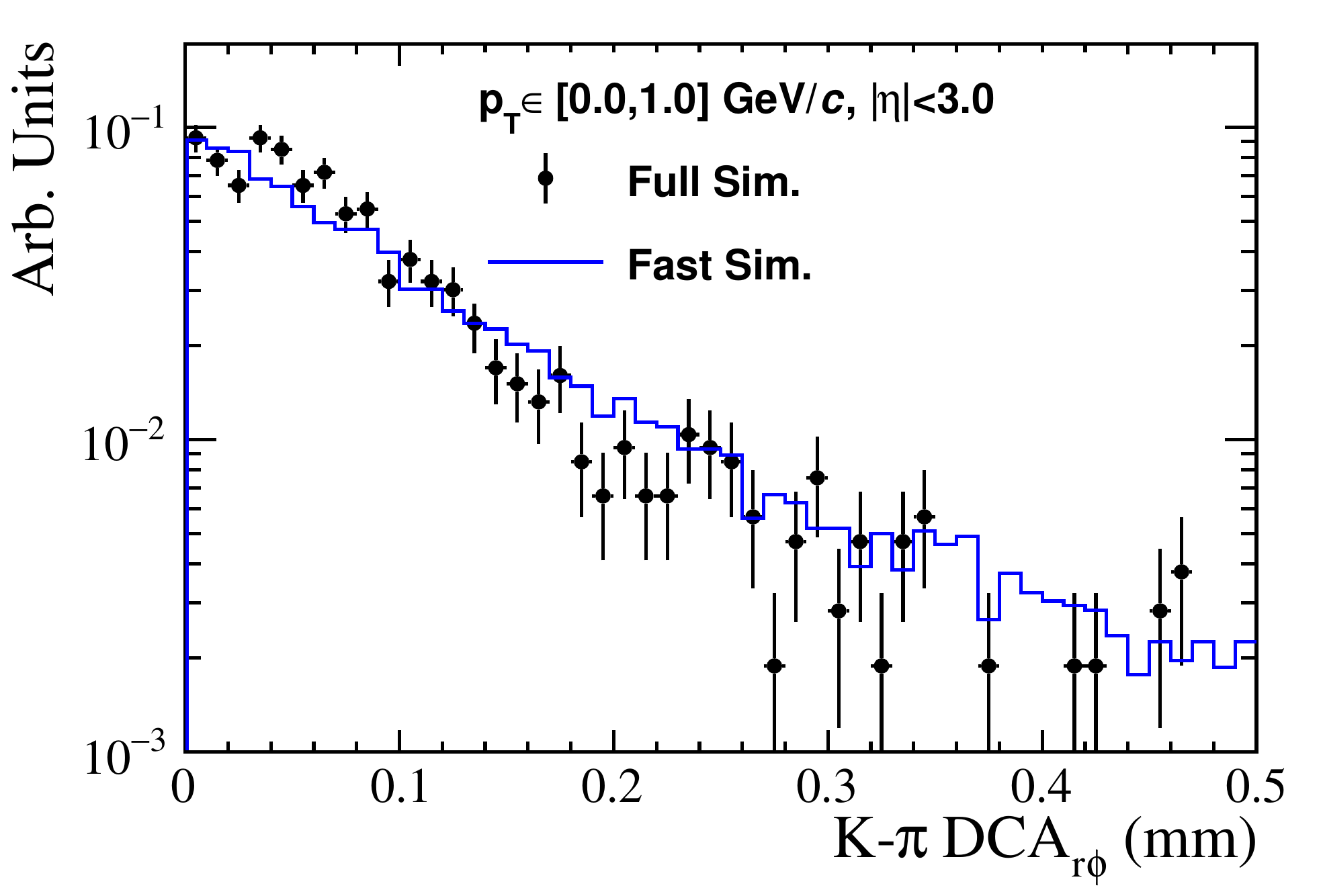}
    \includegraphics[width=0.3\textwidth]{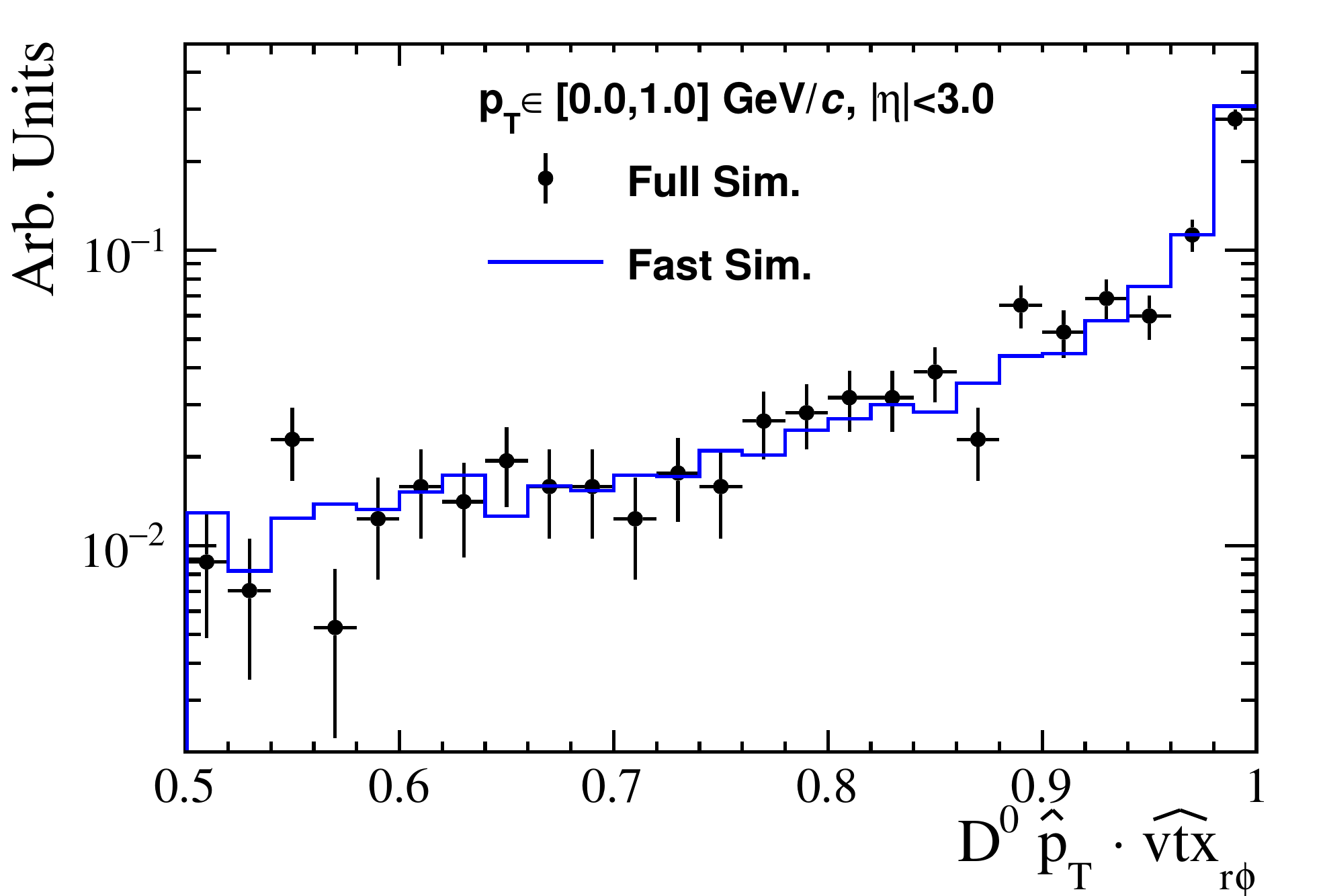}
    \caption{Distributions for $D^{0}$ radial decay length (left), DCA between kaon and pion tracks (middle), and cosine of the $D^{0}$ radial decay vertex pointing angle (right) for the full GEANT4-based simulation (black points) and fast simulation (blue histograms).}
    \label{fig:topocomp}
\end{figure*}

As a systematic check on the fast simulation procedure, we perform the same smearing routine using an alternative detector performance that is directly extracted from the particular GEANT-based simulation described above, and compare the topological variables between the fast and full simulations. These results for an example region of kinematic space are shown in Fig.~\ref{fig:topocomp} with the fast simulation shown as the blue histograms and full simulation as the points (similar quantitative comparisons are seen for all kinematic regions). Within the available statistics of the computationally expensive full simulation there is quite good agreement with the fast simulation and we conclude the fast simulation smearing routine is adequate for our projections within the kinematic regions we study.  
    

\section{Projection for Reduced Charm Cross Sections and Structure Functions}\label{sec:s3}

We first provide the projections for $e$+$p$ collisions as a baseline, and here we focus our studies on $\sqrt{s}=63$ and $\sqrt{s}=29$ GeV energies. The reduced charm cross sections are explicitly calculated as
\begin{widetext}
\begin{equation}
\sigma_{r}^{c\bar{c}}(x_B,Q^{2}) = \frac{dN(D^{0}+\overline{D}^{0})/2}{\mathcal{L}\cdot \varepsilon\cdot \mathcal{B}(D^{0}\rightarrow K\pi)\cdot f(c\rightarrow D^{0}) \cdot \text{d}x_B\text{d}Q^{2}}%\\
\times \frac{x_BQ^{4}}{2\pi\alpha^{2}[1+(1-y)^{2}]},
\end{equation}
\end{widetext}
where $y$ is the inelasticity, $\mathcal{L}$ is the integrated luminosity, $\varepsilon$ is the total efficiency (tracking, PID, reconstruction, and acceptance),  $\mathcal{B}(D^{0}\rightarrow K\pi)$ is the $D^{0}$ branching ratio to $K\pi$, and $f(c\rightarrow D^{0})$ is the $D^{0}$ fragmentation fraction in PYTHIA (56.6\%). As can be observed from the latter quantity, for the purposes of these calculations we scale the measured $D^{0}$ yield to get the total charm cross section. We choose a binning in log$_{10}$($Q^{2}$) and log$_{10}$($x_B$) that is five equal bins per decade along each dimension. 

\begin{figure}[htbp]
    \centering
    \includegraphics[width=0.48\textwidth]{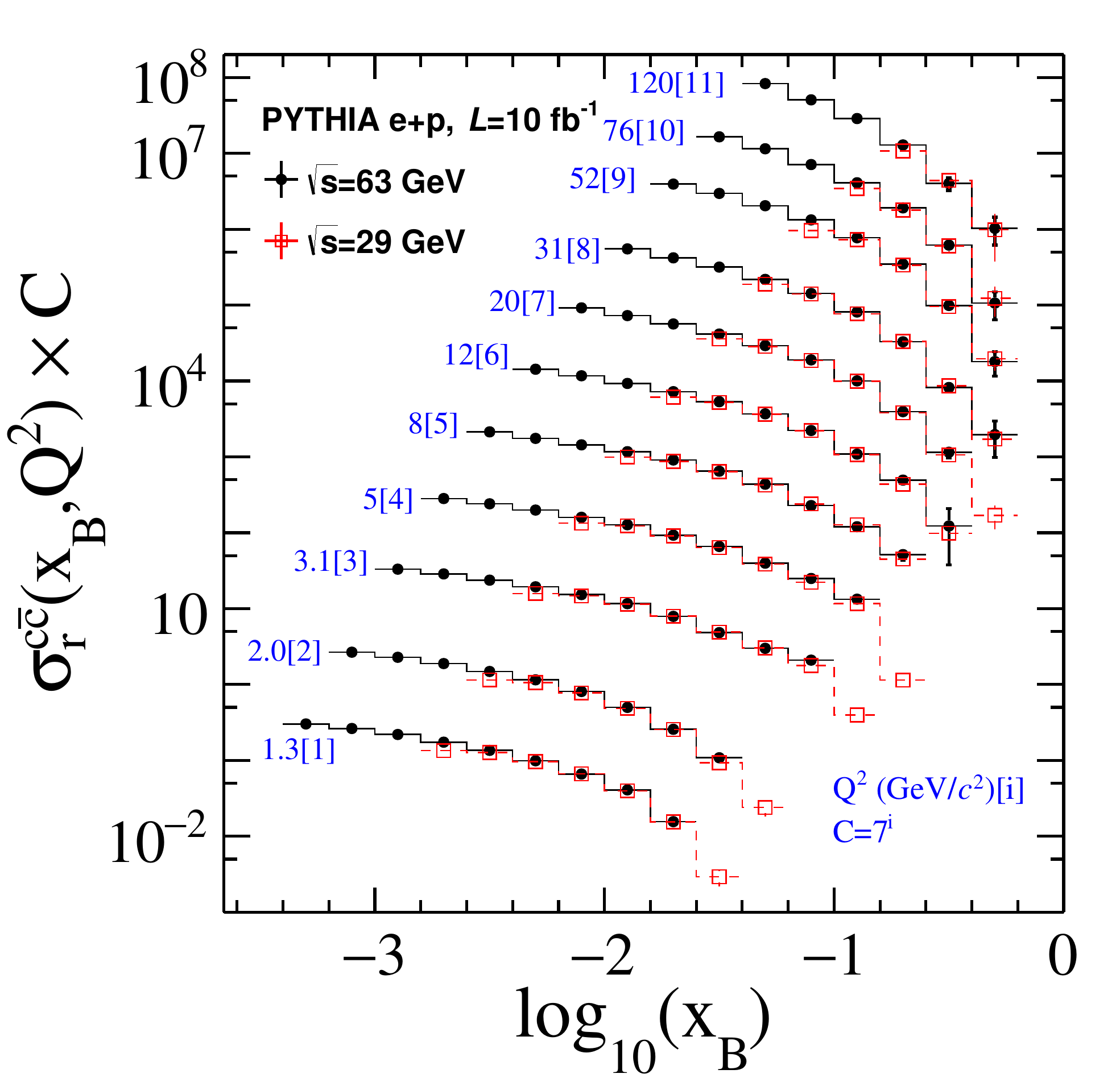}
    \caption{The reduced charm cross section in bins of log$_{10}$($x_B$) and log$_{10}$($Q^{2}$) for $\sqrt{s}=63$ GeV (closed black circles) and $\sqrt{s}=29$ GeV (open red squares) electron+proton collisions in PYTHIA 6 with the charm hadron reconstructed using the all-silicon detector. The vertical values in each $Q^{2}$ bin are scaled by the constant terms $C$ defined in the plot for clarity. The $\sqrt{s}=63$ GeV data is placed at the $x$-axis bin centers while the $\sqrt{s}=29$ GeV is displaced along the $x$-axis for clarity. The statistical uncertainties are scaled to 10 fb$^{-1}$.}
    \label{fig:reducedCS}
\end{figure}

The number of $D^{0}+\overline{D}^{0}$ candidates is determined by counting the number of true $D^{0}\rightarrow K\pi$ decays (post-selection) with invariant mass within $\pm$3$\sigma$ of the peak; The background is calculated in the same fashion and is used to define the signal significance in each bin. We scale all statistical uncertainties to a nominal integrated luminosity of 10 fb$^{-1}$ per center-of-mass energy. The reduced charm cross-sections for $\sqrt{s}=63$ and $\sqrt{s}=29$ GeV $e$+$p$ collisions in bins of $Q^{2}$ and $x_{B}$ are shown in Fig.~\ref{fig:reducedCS} and the uncertainties shown are purely statistical. As can be seen there will be particularly good $Q^{2}$ and $x_{B}$ overlap between the two center-of-mass energies.

\begin{figure}
    \centering
    \includegraphics[width=0.48\textwidth]{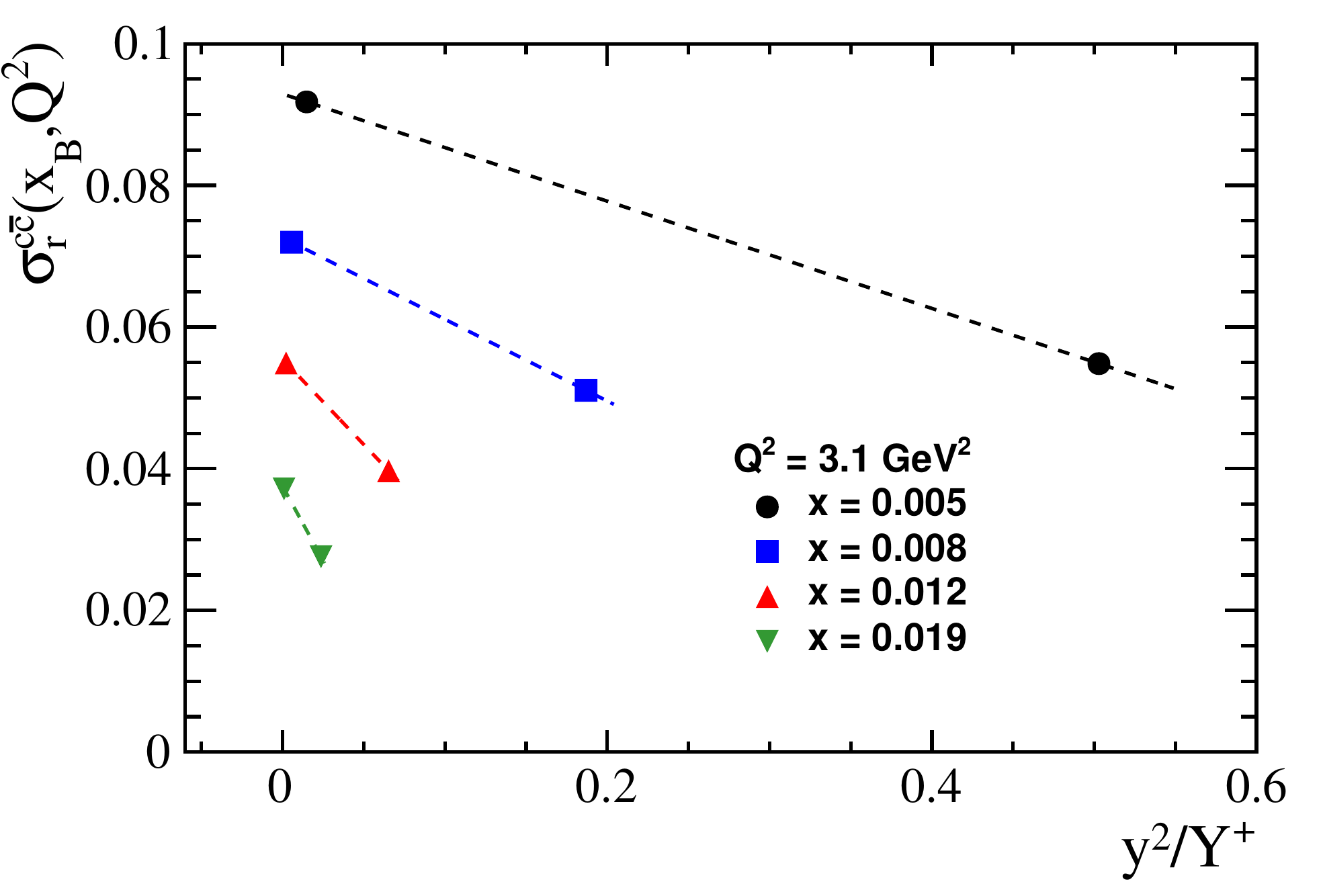}
  
    \caption{Example linear fits to the reduced charm cross sections versus $y^{2}/Y^{+}$ in four different slices of $Q^{2}$ and $x_B$.}
    \label{fig:Charmf2fit}
\end{figure}

\begin{figure}
    \centering
    \includegraphics[width=0.48\textwidth]{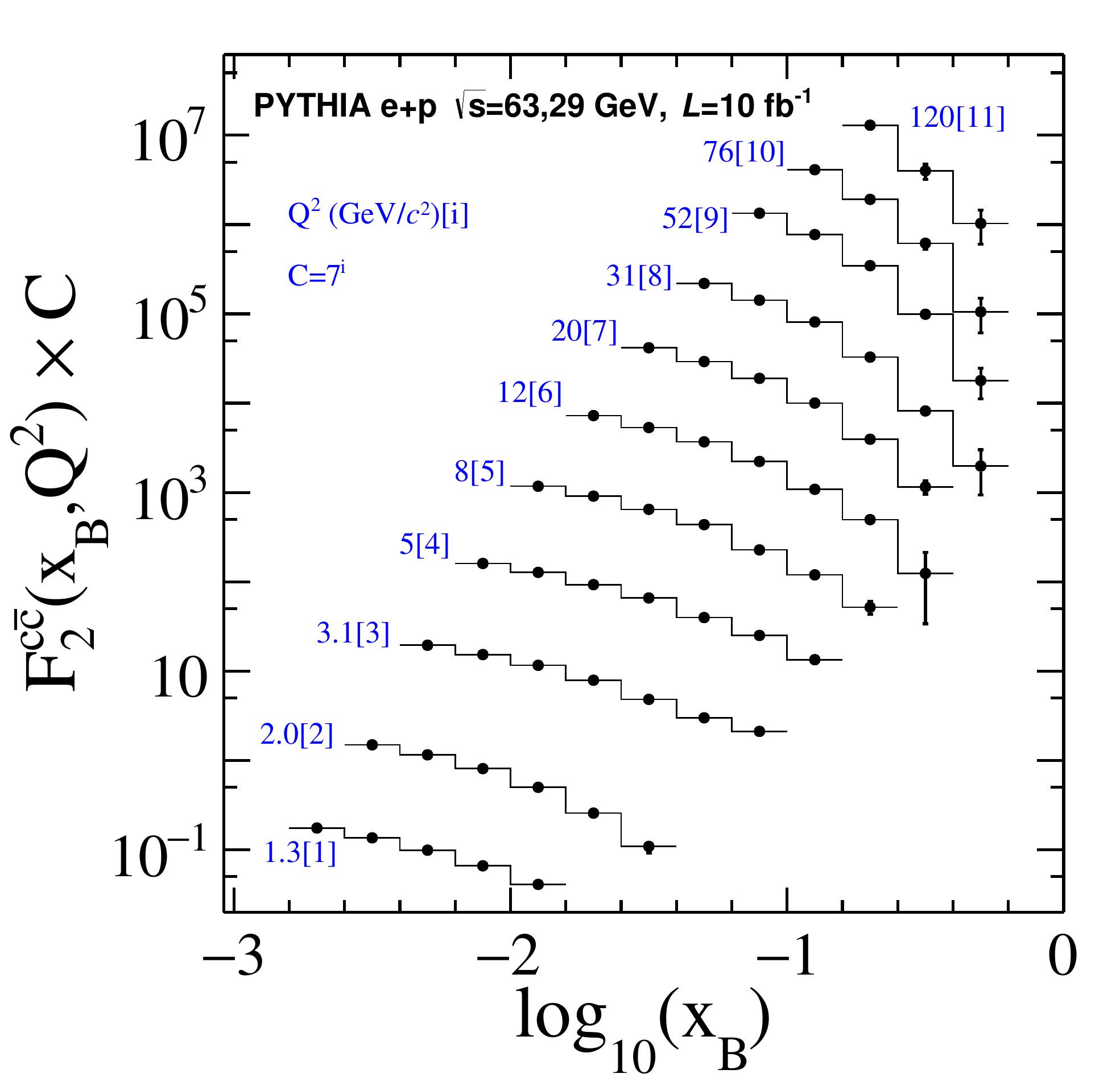}
    \caption{The projected $F_{2}^{c\bar{c}}$ in bins of log$_{10}$($x_B$) and $Q^{2}$. The data points in each $Q^2$ bin are scaled by a factor of $C$ for clarity. The statistical uncertainties are scaled to 10 fb$^{-1}$.}
    \label{fig:Charmf2}
\end{figure}

To extract the charm structure function $F_{2}^{c\bar{c}}$ we use the Rosenbluth technique~\cite{PhysRev.79.615} and take the cross sections at $\sqrt{s}=63$ and $\sqrt{s}=29$ GeV at fixed $x_B$ and $Q^{2}$ and fit the linear form
\begin{equation}
\sigma_{r}^{c\bar{c}}(x_B,Q^{2}) = F_{2}^{c\bar{c}}(x_B,Q^{2}) - \frac{y^{2}}{Y^{+}}F_{L}^{c\bar{c}}(x_B,Q^{2}),
\end{equation}
where $Y^{+} \equiv 1+(1-y)^{2}$. An example fit for four slices of $Q^{2}$ and $x_B$ are shown in Fig.~\ref{fig:Charmf2fit}. The extracted $F_{2}^{c\bar{c}}$ central values and uncertainties from the fits are shown in Fig.~\ref{fig:Charmf2}.


Compared to the work in Refs.~\cite{Aschenauer:2017oxs,Chudakov:2016ytj}, our simulation studies represent a more realistic description of charm-reconstruction capabilities with the EIC detector as we have included PID, momentum and single track pointing resolutions, and primary vertex resolution guided by ongoing detector development/requirements and a full GEANT-based simulation. Furthermore, we have included for the first time the primary vertex resolution in the topological reconstruction of $D^{0}\rightarrow K\pi$ decays in an EIC simulation. 

In Ref.~\cite{Aschenauer:2017oxs} the longitudinal charm structure functions $F_{L}^{c\overline{c}}$ are derived from simulation using charm events tagged by the identification of a displaced kaon vertex, and contain background levels that are less than 2\%. Comparing the kinematic coverage in $Q^{2}$ and $x_{B}$ of $F_{L}^{c\overline{c}}$, our derived $F_{2}^{c\overline{c}}$ has slightly better coverage, particularly in the high-$x_B$ region ($>$0.1). This difference is likely driven by the choice of beam energies used in the simulations and also from the need for three energies to reasonably constrain $F_{L}^{c\overline{c}}$, as opposed to two energies used for $F_{2}^{c\overline{c}}$. However, we note that in Ref.~\cite{Aschenauer:2017oxs} single track and primary vertex resolutions are not folded into the kaon distributions. Incorporating these resolutions could significantly smear the charm and background kaon vertex distributions, and consequentially reduce the charm event purity and limit the kinematic coverage. Therefore, in corroboration with Ref.~\cite{Aschenauer:2017oxs}, our studies show that with a detector response measurements of charm structure functions will be possible across a broad kinematic range at the EIC. We also note that a precise measurement of $F_{L}^{c\overline{c}}$ will still be feasible with a full detector response, but as shown in Fig.~\ref{fig:cq2x} with the needed three center-of-mass energies the overlap in $Q^{2}$ and $x_{B}$ will be limited.

\section{Impact of Intrinsic Charm in the Proton}\label{sec:s4}

To estimate the impact on the charm structure functions $F_{2}^{c\overline{c}}$ with the existence of IC, we replace the default PDFs in PYTHIA 6.4 with those from the recent CT14 analysis~\cite{Hou2018} with and without IC components. There are several IC models used in the CT14 analysis (and hence different PDF sets), and for our studies we pick two which encapsulate two valiant features of the different models: A scenario in which the IC is "sea-like"(Sea1) and another as "Valence-like"(BHPS1). Both Sea1 and BHPS1 IC models are described in more detail and compared to data in~\cite{Hou2018}. The former sea-like model has an IC PDF that is spread across all values of $x_{p}$, while the latter valence-like produces an IC PDF that is concentrated at values of $x_{p}>$0.1. In both models, the average total momentum fraction of the IC is on the order of 1\%. 

\begin{figure}
    \centering
    \includegraphics[width=0.48\textwidth]{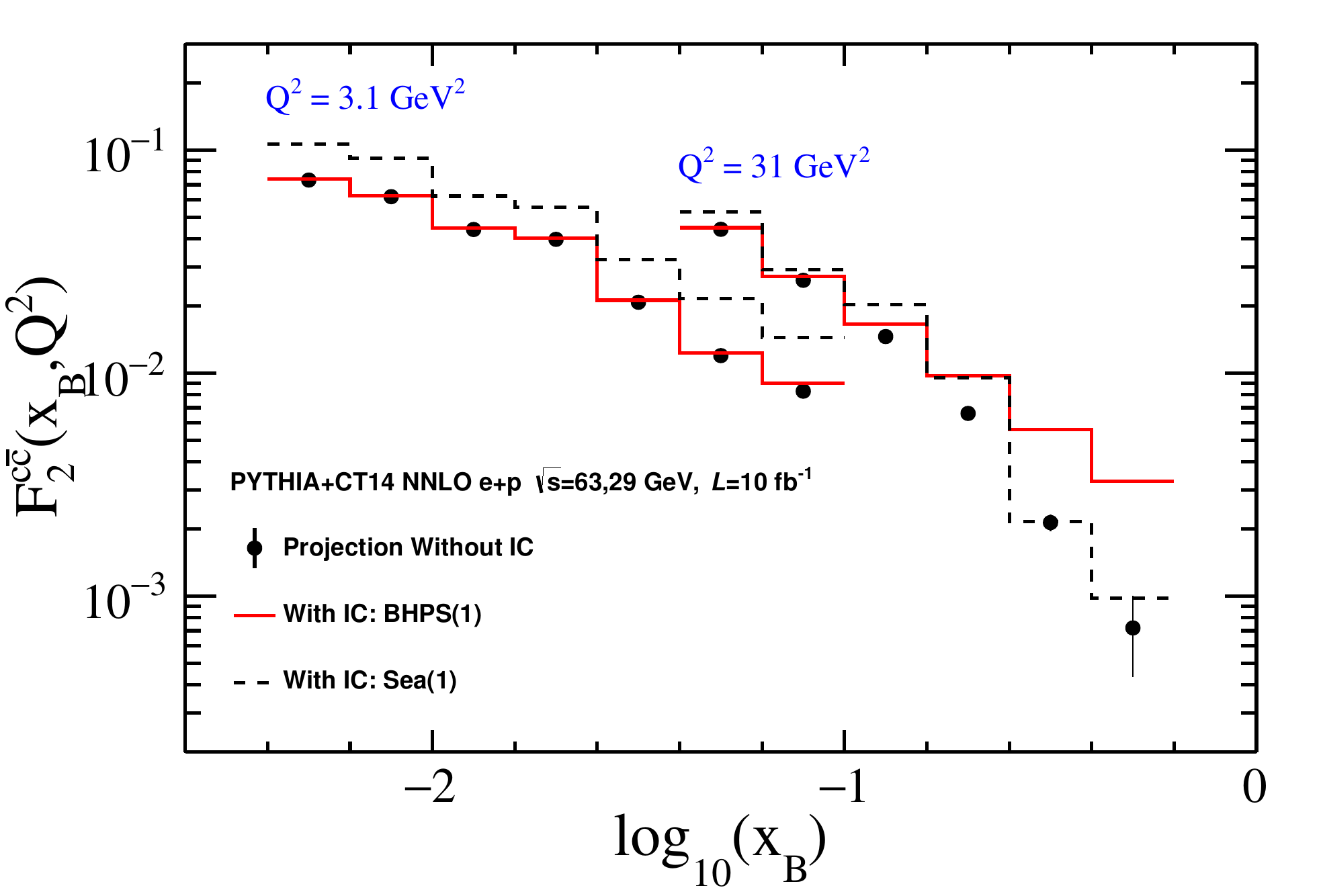}
    \caption{The projected $F_{2}^{c\overline{c}}$ with the baseline CT14 PDF (black points), CT14 with sea-like intrinsic charm (dashed black lines), and CT14 with BHPS-like intrinsic charm (red lines), for two representative values of $Q^{2}$. The statistical uncertainties are scaled to 10 fb$^{-1}$.}
    \label{fig:Charmf2IC_rep}
\end{figure}

\begin{figure}
    \centering
    \includegraphics[width=0.48\textwidth]{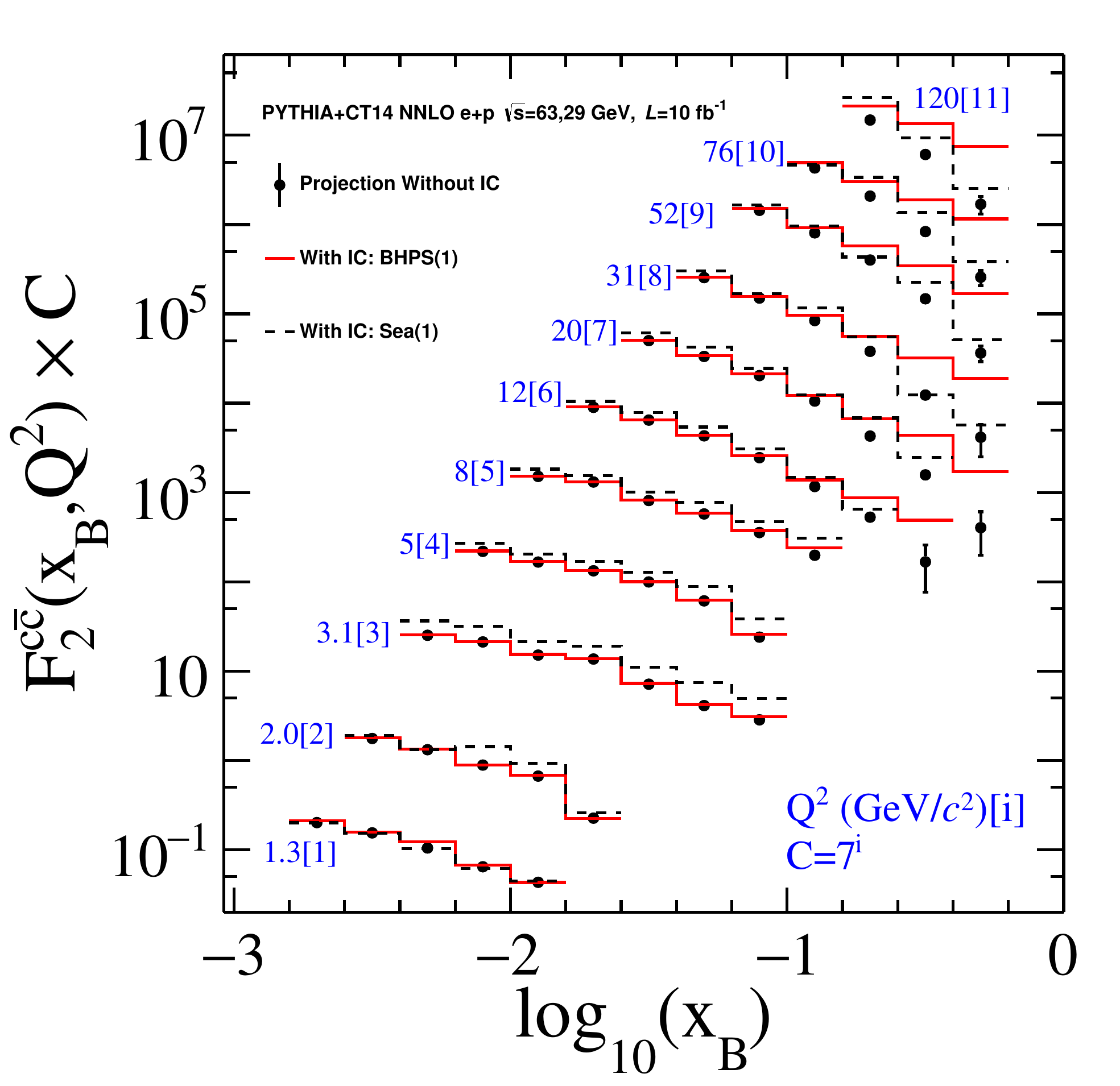}
    \caption{The projected $F_{2}^{c\overline{c}}$ with the baseline CT14 PDF (black points), CT14 with sea-like intrinsic charm (dashed black lines), and CT14 with BHPS-like intrinsic charm (red lines). The data points in each $Q^2$ bin are scaled by a factor of $C$ for clarity. The statistical uncertainties are scaled to 10 fb$^{-1}$.}
    \label{fig:Charmf2IC}
\end{figure}

The effect of including IC in the CT14 PDFs on our $F_{2}^{c\overline{c}}$ projections can be seen in Fig.~\ref{fig:Charmf2IC_rep} for two representative bins of $Q^{2}$, and in Fig.~\ref{fig:Charmf2IC} across all bins of $Q^{2}$ and $x_{B}$. Here the black data points correspond to our projections without any IC, and the black dashed line and red solid line show the enhancement from including a sea-like and valance-like IC PDF, respectively. As expected the sea-like model produces an enhancement across all values of $x_{B}$ as the IC PDF is finite across a broad $x_{p}$ range, and is about a factor of two larger with respect to the projection without IC. The valence-like model produces a significant enhancement at $x_{B}>$0.1, which is about a factor of seven at very large $x_{B}$. No enhancement is observed at lower values of $Q^{2}$ for the valence-like model because in this region the data do not probe values of $x_{p}$ where the valence-like IC PDF is significant. 

We have also repeated this exercise with the NNPDF3 PDF set\footnote{We explicitly use the NNPDF3\_IC\_nlo\_as\_0118\_mcpole\_1470 and NNPDF3\_nIC\_nlo\_as\_0118\_mcpole\_1470 PDF sets provided by the NNPDF collaboration} where the intrinsic charm content of the proton is determined using a "fitted" charm method~\cite{Ball2016}. The IC PDF from the NNPDF analyses is comparable to the BHPS1 model in the CT14 analysis and peaks at slightly large values of $x_{p}$, and has a total momentum fraction also on the order of 1\%. We observe that the NNPDF3 IC produces an enhancement in the projected data for $x_{B}>$0.1 that is qualitatively similar to the CT14 valence-like model. The absolute enhancement is a factor of 10 for $x_{B}>$0.1 (compared to a factor of seven for the CT14 BHPS1 model). We find no enhancement in the projected data with NNPDF3 PDFs below a $Q^{2}$ value of 8 GeV$^{2}$ where the data probe values of $x_{B}<$0.1.

With both IC models the observed enhancement with respect to the no-IC scenario is well above the projected statistical uncertainties of the base-line data. These studies show at the EIC measurements of charm structure functions will be extremely sensitive to the existence of IC and to different IC models.

\section{Constraints to the Nuclear Gluon Parton Distribution Functions}\label{sec:s5}

\begin{figure*}
    \centering
    \includegraphics[width=0.65\textwidth]{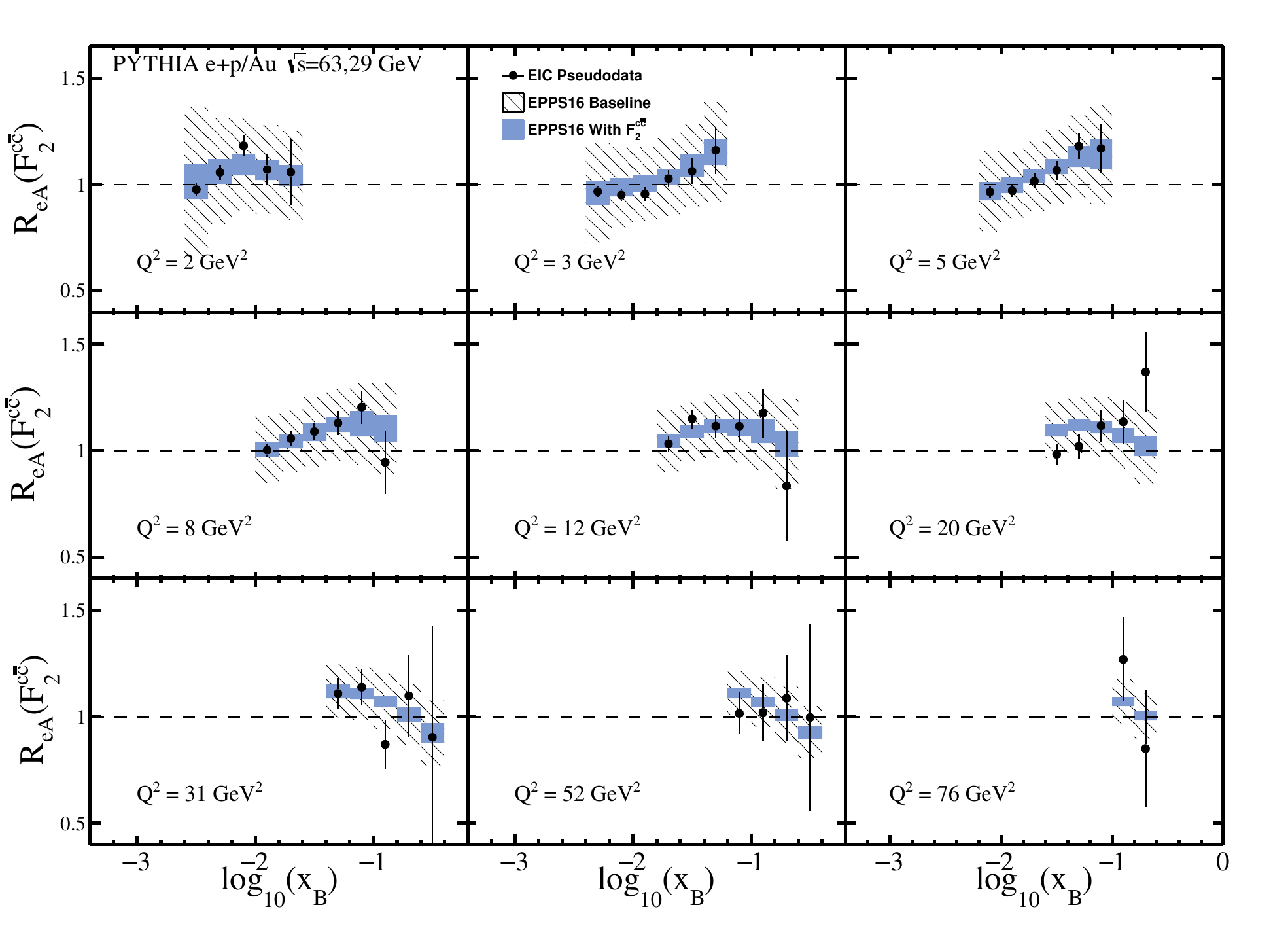}
    \caption{The ratios of $F_{2}^{c\overline{c}}$ in $e$+Au to $e$+$p$ collisions in bins of $Q^{2}$ and $x_{B}$ with integrated luminosities of 1 fb$^{-1}$/nucleon. The hashed gray curves show the EPPS16~\cite{Eskola:2016oht} baseline uncertainties while the solid blue curves show the re-weighted uncertainties. }
    \label{fig:F2Ratios}
\end{figure*}

To project the statistical precision of a charm structure function measurement in $e$+Au collisions we utilize the CT14+EPPS16 Au PDFs~\cite{PhysRevD.93.033006,Eskola:2016oht} in our PYTHIA simulation and repeat the same procedure outline in Section~\ref{sec:s3}. CT14+EPPS16 is used as an example here but we also repeat the exercise for nCTEQ15~\cite{Kovarik:2015cma} and nNNPDF2.0~\cite{Khalek_2020} PDFs for Au nuclei (The respective proton baseline PDFs used are CT14~\cite{PhysRevD.93.033006} and NNPDF3.1~\cite{Ball_2017} of the same order in the strong coupling constant and with the same value of the strong coupling constant). The projected ratios of $F_{2}^{c\overline{c}}$ in $e$+Au collisions with respect to $e$+$p$ are shown in Fig.~\ref{fig:F2Ratios} as the black data points. The data here have been randomly displaced from the central value using the statistical uncertainties in both $e$+Au and $e$+$p$ collisions, and we note the total integrated luminosity per energy and colliding system here is 1 fb$^{-1}$/nucleon. The gray hashed curves show the uncertainty on the $F_{2}^{c\overline{c}}$ ratio coming from the EPPS16 gluon\footnote{Our goal is only to study the impact on the nuclear gluon PDF and we are not performing a complete QCD fit, therefore we only select charm hadrons in PYTHIA that originate from the photon-gluon fusion process to be used in the Bayesian PDF re-weighting procedure. We then weight this sub-sample of data to conserve the total yield and uncertainty from all processes in each bin of $Q^{2}$ and $x_{B}$.} nPDFs uncertainties only, and are in general much larger than the projected data uncertainties.

To estimate the impact on the nuclear gluon PDF uncertainties we utilize the Bayesian PDF re-weighting procedure~\cite{Paukkunen2014}. In the case of EPPS16 and nCTEQ15 PDFs, where the PDF errors are Hessian, we generate a large sample of PDF replicas using
\begin{equation}\label{eq:rep}
f_{k} = f_{0} + \sum_{i}\Big(\frac{f_{i,+} + f_{i,-}}{2}\Big)r_{k,i},
\end{equation}
where $f_{0}$ is the PDF central value, the index $i$ runs over all eigen PDF error sets, and $f_{i,+/-}$ are the deviations from the central value of the $i$-th plus/minus PDF error. $r_{k,i}$ is a random Gaussian number centered at zero with unit variance. The re-weighted (Hessian) PDFs are then constructed by 
\begin{equation}\label{eq:newrep}
f_{new} = f_{0} + \sum_{i}\Big(\frac{f_{i,+} + f_{i,-}}{2}\Big)\Big[\frac{1}{N_{rep}}\sum_{k}w_{k}r_{k,i}\Big],
\end{equation}
with the index $k$ running over all replicas with weights $w_{k}$. Here we choose the weights proposed by Giele and Kelle~\cite{PhysRevD.58.094023}
\begin{equation}\label{eq;gk}
w_{k} = \frac{exp[-\chi_{k}^{2}/2]}{\frac{1}{N_{rep}}\sum_{k}exp[-\chi_{k}^{2}/2]}.
\end{equation}

In the case of the nNNPDF2.0 PDFs we use the set of already produced Monte Carlo replicas and the weights advocated for by the NNPDF collaboration~\cite{Ball:2010gb,Ball:2011gg}
\begin{equation}\label{eq:nnpdf}
w_{k} = \frac{(\chi_{k}^{2})^{\frac{1}{2}(n-1)}exp[-\chi_{k}^{2}/2]}{\frac{1}{N_{rep}}\sum_{k}(\chi_{k}^{2})^{\frac{1}{2}(n-1)}exp[-\chi_{k}^{2}/2]},
\end{equation}
where $n$ is the number of data points in the fit.

\begin{figure*}[!htbp]
    \centering
    \includegraphics[width=0.8\textwidth]{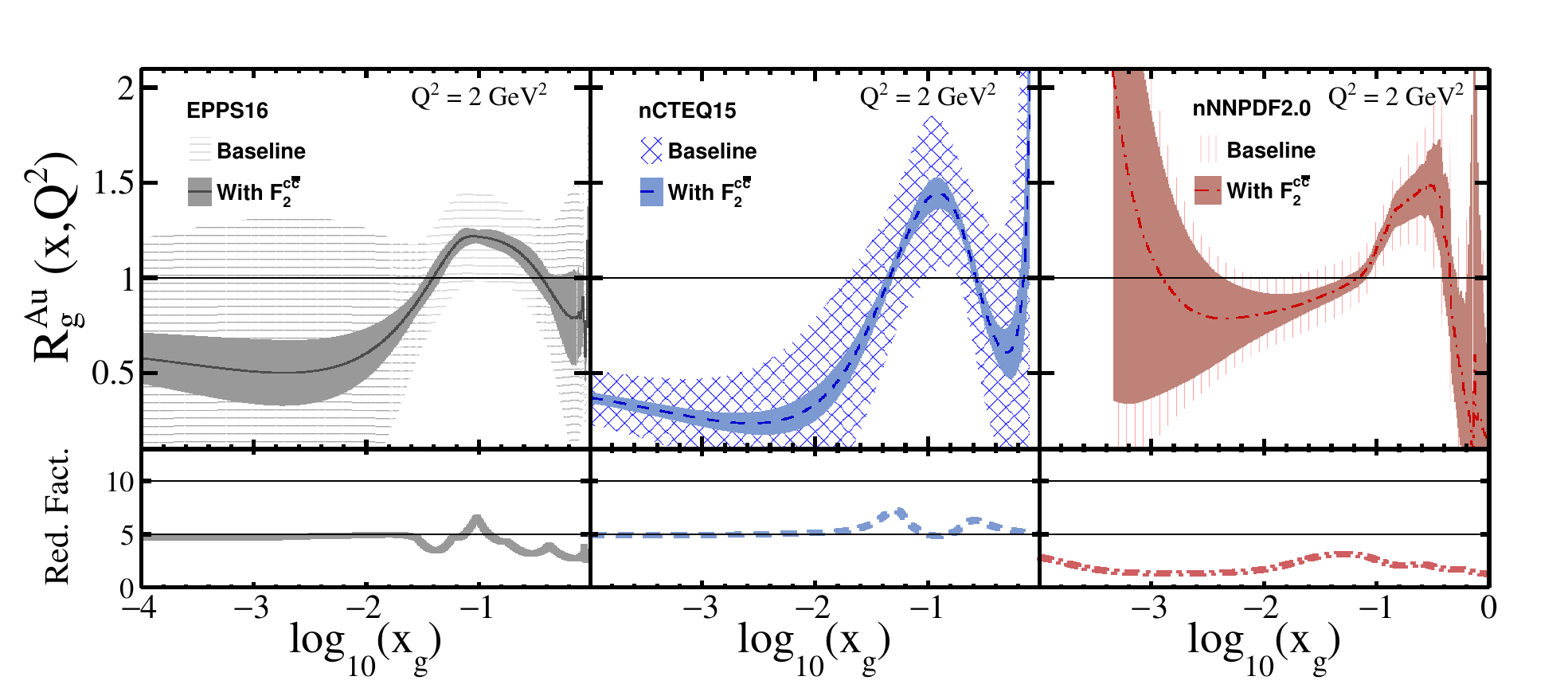}
    \caption{Top: The nuclear gluon ratios as a function of $x_g$ for Au nuclei for the EPPS16~\cite{Eskola:2016oht} (left), nCTEQ15~\cite{Kovarik:2015cma} (middle), and nNNPDF2.0~\cite{Khalek_2020} (right) PDF sets. The hashed curves show the baseline uncertainties and solid curves show the re-weighted uncertainties using the projected charm data with integrated luminosities of 1 fb$^{-1}$/nucleon. Bottom: The reduction factor of the nPDF uncertainties with the inclusion of the EIC charm data.}
    \label{fig:Gluon1}
\end{figure*}

\begin{figure*}[!htbp]
    \centering
    \includegraphics[width=0.8\textwidth]{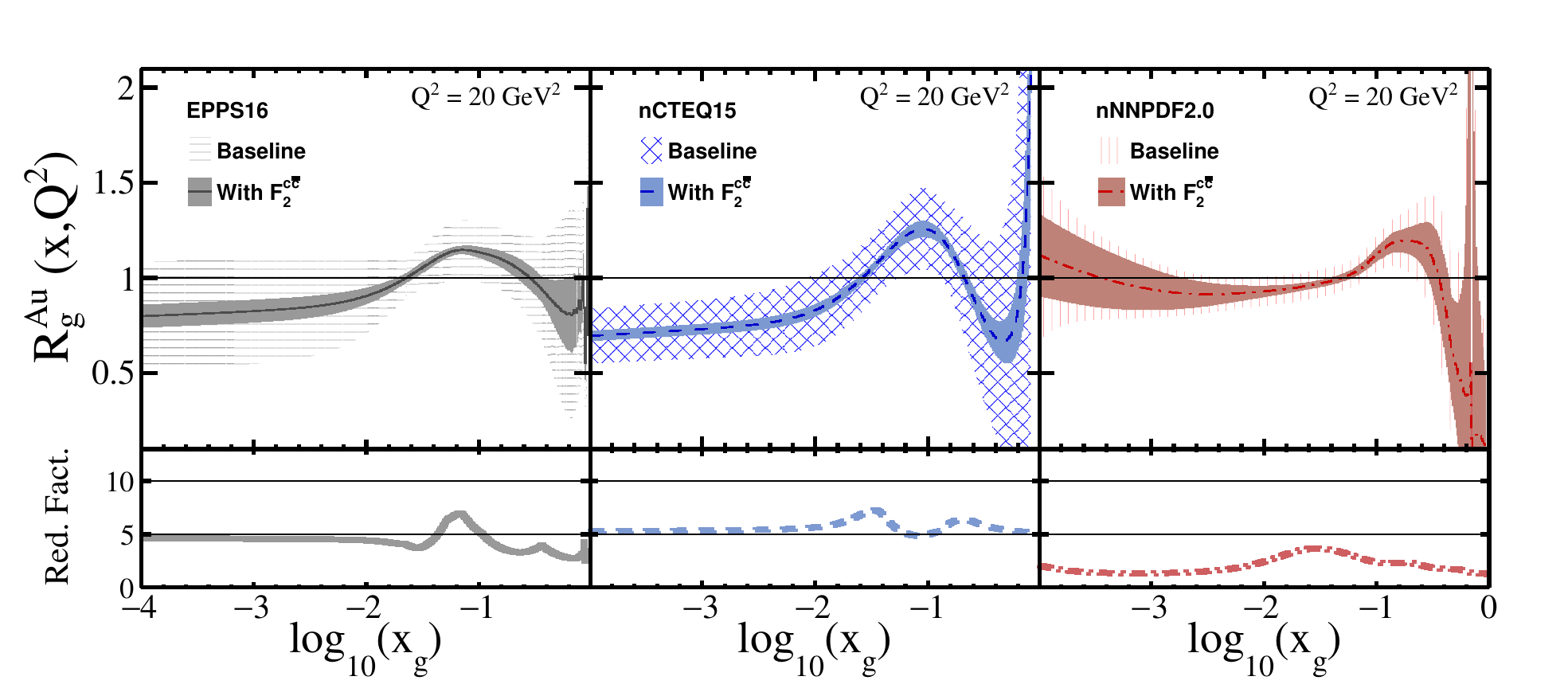}
    \caption{Same as Fig.~\ref{fig:Gluon1} but for $Q^{2}=20$ GeV$^{2}$.}
    \label{fig:Gluon2}
\end{figure*}

\begin{figure*}[!htbp]
    \centering
    \includegraphics[width=0.8\textwidth]{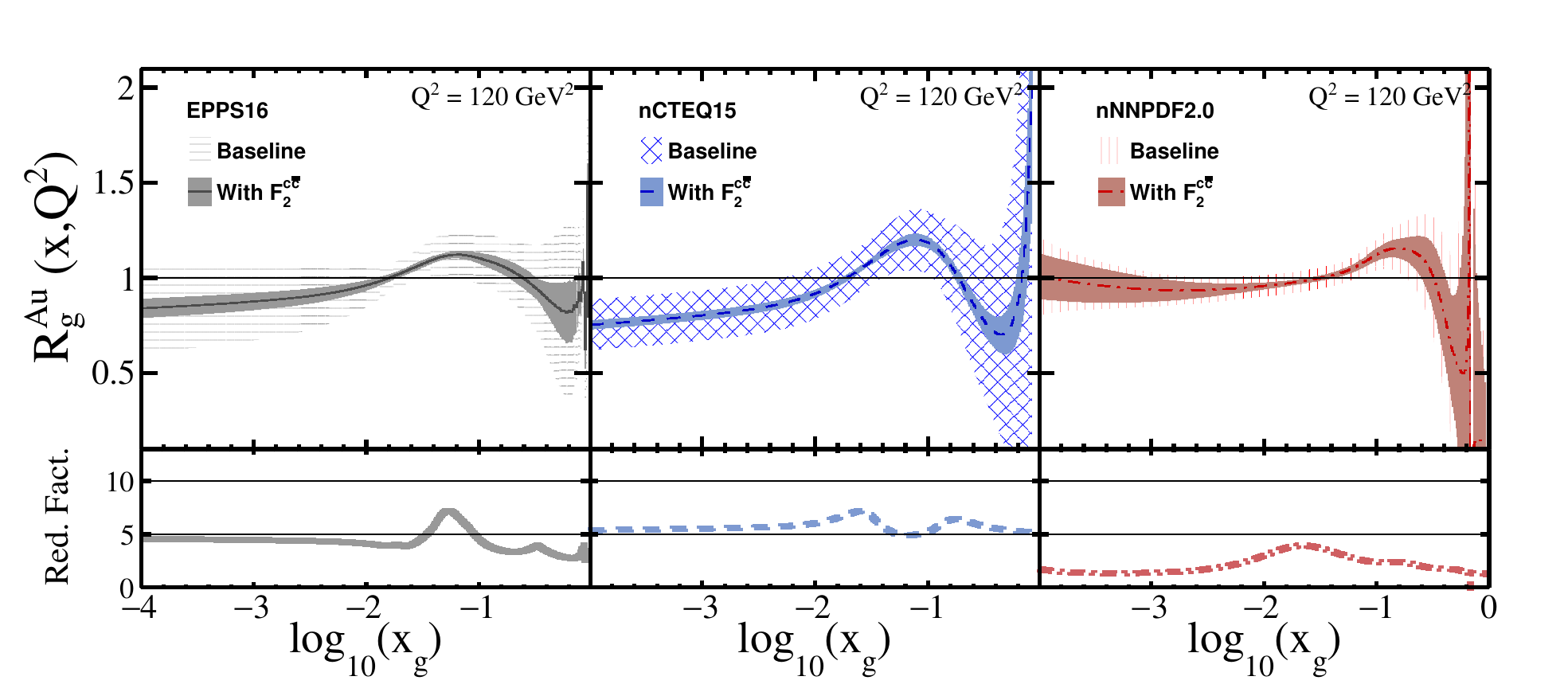}
    \caption{Same as Fig.~\ref{fig:Gluon1} but for $Q^{2}=120$ GeV$^{2}$.}
    \label{fig:Gluon3}
\end{figure*}

The EPPS16, nCTEQ15, and nNNPDF2.0 nuclear gluon ratios before and after re-weighting are show in Fig.~\ref{fig:Gluon1} for $Q^{2}$ = 2 GeV$^{2}$; Figs.~\ref{fig:Gluon2}~and~\ref{fig:Gluon3} show the same for $Q^{2}$ = 20 GeV$^{2}$ and $Q^{2}$ = 120 GeV$^{2}$, respectively. The improvement of the gluon nPDF uncertainties, shown in the bottom panels, is estimated to be about a factor of five for the EPPS16 and nCTEQ15 nPDFs across all values of gluon momentum fraction. For the nNNPDF2.0 gluon nPDF, the improvement is on average about a factor of two, four, and three at values of $x_{g}<0.01$, $x_{g}\thicksim 0.01-0.1$, and $x_{g}>0.1$, respectively. The re-weighted EPPS16 uncertainties for the predicted $F_{2}^{c\overline{c}}$ ratios are shown in Fig.~\ref{fig:F2Ratios} as the shaded blue bands. 


We have also studied the effect of using different weights for the EPPS16 and nCTEQ15 re-weighting procedure. We have first performed the analyses with the weights advocated for by the NNPDF collaboration shown in Eq.~\ref{eq:nnpdf}. In general, we find these weights produce a comparable reduction in the gluon nPDF uncertainties compared to the nominal weights. It has been argued that the Giele-Keller weights should be scaled by taking $\chi^{2}_{k}\rightarrow \chi^{2}_{k}/\Delta\chi^{2}$ in Eq.~\ref{eq;gk}, which takes into account the tolerance $\Delta\chi^{2}$ criteria used in the PDF error analyses~\cite{Paukkunen2014}. Scaling the Giele-Keller weights in this way broadens the weight distribution versus $\chi^{2}_{k}$. We have also repeated the analysis with the scaled weights and find the reduction in the nPDF uncertainties is improved rather than deteriorated. We attribute this to the fact that the $\chi_{k}^{2}$ distribution normalized by the number of degrees of freedom ($n.d.f.$) has long tails at large values, and for example EPPS16 has $\langle\chi_{k}^{2}\rangle/n.d.f.$= 19. In effect, this scaled weighting procedure includes more information utilizing a significant fraction of replicas that have large deviation with respect to the data, hence producing an improvement in the re-weighted nPDF uncertainties. However, this highlights that the new pseudo-data have more constraint than existing data used the global fits potentially indicating a full re-fit of the data is needed. Therefore, we consider our improved constraints shown in Figs.~\ref{fig:Gluon1},~\ref{fig:Gluon2}, and~\ref{fig:Gluon3} as a lower limit.

It has been shown in Ref.~\cite{Aschenauer:2017oxs} that the low-$x_{g}$ region will already be well constrained from inclusive measurements, with little improvement from charm measurements. Our studies are in corroboration with those in Ref.~\cite{Aschenauer:2017oxs}, which use a more flexible EPPS16* nPDF set (particularly at regions of partonic $x_{p}$ less than the anti-shadowing peak), and show the high-$x_{g}$ region will be significantly constrained from open charm measurements. Therefore, confirming that the inclusive and charm measurements will be complementary to constraining the nuclear gluon PDF across all $x_{g}$. 

\section{Conclusions}\label{sec:s6}
To summarize, we have studied the estimated precision of charm hadron production measurements using a fast simulation that includes detector performance from ongoing detector development. Using the exclusive $D^{0}\rightarrow K^{-}\pi^{+}$ decay channel we have estimated the expected statistical precision of the reduced charm cross sections in $\sqrt{s}=63$ and $\sqrt{s}=29$ GeV e+p collisions, and $F_{2}^{c\overline{c}}$ in a data set with an integrated luminosity of 10 fb$^{-1}$ per energy. With the detector resolutions used we show the identification of the charm hadron decay vertex will significantly reduce the backgrounds in a EIC environment. This will improve the signal significance and reduce systematic uncertainties associated with the signal extraction.

We studied the effect of using proton PDFs that include several IC models on the observed $F_{2}^{c\overline{c}}$ in PYTHIA and find $F_{2}^{c\overline{c}}$ is enhanced well above the projected statistical uncertainties and those coming from the baseline PDFs. For the IC PDFs determined in the CT14 BHPS1 model and NNPDF3 fitted charm, this enhancement is found to be nearly an order of magnitude at $x_{p}>0.1$. These data will be important for understanding the proton charm PDFs and for an accurate determination of all the proton PDFs. 

We have also studied the constraint on the gluon nPDFs coming from the EIC charm measurements using a Bayesian PDF re-weighting procedure. We have incorporated several nuclear PDFs in our simulation, and have calculated the ratio of $F_{2}^{c\overline{c}}$ in $e$+Au to $e$+$p$ with 1 fb$^{-1}$/nucleon worth of data per energy and colliding system.  Incorporating the EIC charm data in a Bayesian PDF re-weighting procedure shows that relative gluon nPDF uncertainties can be reduced by a factor of about 3 to 7 depending on the nPDF used and kinematic region. This will be particularly important in the high-$x_{p}$ region where inclusive measurements will have little constraint, thus highlighting the necessity of these measurements.

\section*{Acknowledgements}\label{sec:s7}
We thank Feng Yuan for invaluable discussions throughout the course of these studies. We also thank the RCF facility at Brookhaven National Laboratory and the NERSC center and Lawrence Berkeley National Laboratory for the computational resources used for this work. This work was supported in part by the Office of Nuclear Physics within the U.S. DOE Office of Science and the U.S. National Science Foundation.

\bibliography{bib}

\end{document}